\documentclass[aps,prl,twocolumn,showpacs,preprintnumbers,longbibliography,amsmath,amssymb,floatfix,superscriptaddress]{revtex4-2}
\usepackage{graphicx}
\usepackage{amsfonts}
\usepackage{CJK}
\usepackage{bm}
\usepackage{color}
\usepackage{xcolor}
\usepackage{array}
\usepackage[T1]{fontenc}
\usepackage[colorlinks = true,linkcolor = blue,urlcolor = blue,citecolor = blue,anchorcolor = blue]{hyperref}

\begin{document}

\title{Approaching the Landauer Bound without Slowing Down}
\author{Yue Liu}\thanks{These authors contributed equally to this work.}

\affiliation{Department of Physics and Shenzhen Research Institute, Xiamen University, Xiamen 361005, Fujian, China }

\affiliation{Center for Gravitational Physics and Quantum Information, Yukawa Institute for Theoretical Physics, Kyoto University, Kitashirakawa Oiwakecho, Sakyo-ku, Kyoto 606-8502, Japan}

\author{Chenlong Huang}\thanks{These authors contributed equally to this work.}

\affiliation{Department of Physics and Shenzhen Research Institute, Xiamen University, Xiamen 361005, Fujian, China }

\author{Xingyu Zhang}

\affiliation{Department of Physics and Shenzhen Research Institute, Xiamen University, Xiamen 361005, Fujian, China }

\affiliation{School of Science, Zhejiang University of Science and Technology, Hangzhou 310023, China}

\author{Dahai He}\email{dhe@xmu.edu.cn}

\affiliation{Department of Physics and Shenzhen Research Institute, Xiamen University, Xiamen 361005, Fujian, China }

\date{\today}
\begin{abstract}
  Finite-time memory erasure is conventionally understood to incur an additional energy cost beyond the Landauer bound. Whether this bound can be approached without taking the infinite-time limit remains open. Based on a quantum speed limit, we identify coherence between degenerate states as a resource that allows the Landauer bound to be approached at a finite operation time. For both the $2N$-state and collective spin models, we show that the minimum dissipated heat and the heat generated by continuous protocols approach the Landauer bound at finite operation times, with the corresponding finite-time penalty scaling inversely with system size. We further validate these theoretical predictions numerically for systems coupled to the Ohmic bosonic heat bath with various driving protocols. Our results reveal that operation time is a substitutable resource for approaching the fundamental thermodynamic bound of information erasure.

\end{abstract}
\maketitle
\textit{Introduction.}---Memory erasure is a fundamental thermodynamic task in information processing and quantum computation. The energy cost required to erase a bit of information is dictated by the Landauer principle~\cite{Landauer1961, Bennett1982, Sagawa2012,Parrondo2015, Georgescu2021}
\begin{equation}\label{Landauer}
  Q\geq \beta^{-1}\ln{2},
\end{equation}
where $\beta=1/(k_{\text{B}}T)$ represents the inverse temperature of the environment. The Landauer principle has been experimentally validated across various platforms~\cite{Antoine2012,Jun2014,Hong2016,Yan2018,Gaudenzi2018,Dago2021,Dago2022,Scandi2022,Dago2023}. Analogous to the Carnot efficiency, the Landauer bound is typically attainable only in the quasistatic limit, where processes occur infinitely slowly. However, practical applications demand highly efficient and fast operations~\cite{Magnard2018,Zhou2021,Johnson2022,Reiner2024,Alibaba2024}, which results in the thermodynamic cost being many orders of magnitude above the Landauer bound~\cite{Wolpert2024}. Hence, significant efforts have been dedicated to studying the finite-time Landauer bound~\cite{Zhen2021, Basilewitsch2021, Tan20221, Zhen2022, Lee2022, Tan2023} by various thermodynamic inequalities and trade-off relations~\cite{Peliti2021,Shiraishi2023}, and determining the minimal thermodynamic cost and the optimal protocol of finite-time memory erasure~\cite{Proesmans2020L, Proesmans2020E, Ma2022, Scandi2022, Rolandi2023, Rolandi2023b, Dong2024, Li2024, Liu20253}. 

As devices approach the atomic scale, quantum effects inevitably emerge and play an increasingly pivotal role in thermodynamic processes~\cite{Binder2018,Campbell2026}. Such effects can modify conventional trade-offs in classical finite-time thermodynamics. For example, quantum coherence can alter the relation between power and efficiency in thermal machines~\cite{Scully2010,Scully2011,Brandner2017,Cao2018,Brandner2020,Jaegon2022,Kamimura2022,Mohan2025,Funo2025} and the precision bounds set by thermodynamic uncertainty relations~\cite{Hasegawa2020,Kalaee2021,Tan20222,Xingyu2025,Tan2025,Prech2025}. In particular, quantum coherence and degeneracy can allow Carnot efficiency to be approached at finite or even maximum power~\cite{Tajima2021,Liang2025}. Information erasure has also been studied in the quantum regime~\cite{Timpanaro2020,Miller2020,Tan20221,Tan2023}, with coherence between different energy eigenstates shown to increase the thermodynamic cost of qubit reset~\cite{Miller2020,Tan20221}. A missing piece, however, is whether the Landauer bound can be approached without extending the operation time by trading this valuable practical resource for a scalable quantum resource.

In this Letter, we show that the Landauer bound can be approached at a finite erasure time. A quantum speed limit formulated in terms of the quantum Wasserstein distance identifies coherence between degenerate states as the resource that enables this finite-time approach. By introducing a $2N$-state model with degeneracy $N$ and a collective spin model, we demonstrate that a lower bound on the finite-time thermodynamic cost converges to the Landauer bound as the system size increases. We further investigate the minimal thermodynamic cost and continuous erasure protocols. These results show that the Landauer bound can be approached by scaling up quantum resources rather than extending the operation time.


\textit{Setup.}---Information erasure occurs over a finite time $\tau$ through a controllable open quantum system weakly coupled to a heat bath at inverse temperature $\beta$. The time evolution of the density matrix $\rho$ of the system is governed by the Lindblad master equation
\begin{equation}\label{Lindblad}
  \dot{\rho}=-i\hbar^{-1}\left[H(t),\rho\right]+\mathcal{D}[\rho],
\end{equation}
where $H(t)$ denotes the time-dependent Hamiltonian spanned by energy eigenstates $|c_{j}\rangle$ with energy eigenvalues $c$ and degeneracy index $j$, and $\mathcal{D}[\rho]:=\sum_{k}(L_{k}\rho L_{k}^{\dagger}-\{L_{k}^{\dagger}L_{k},\rho\}/2)$ represents the dissipative term. The jump operator $L_{k}$ describes the transition between energy eigenstates with energy gap $\omega_{k}$, satisfying $[L_{k},H(t)]=\omega_{k} L_{k}$. Hereinafter, the Planck constant and the Boltzmann constant are set to unity ($\hbar=k_{\text{B}}=1$). We assume the local detailed balance condition $L_{k}=e^{\beta\omega_{k}/2}L^{\dagger}_{k'}$, where the jump operator $L_{k'}$ describes the reverse jump process of $L_{k}$.

Information is encoded in the density matrix $\rho(t)$ and quantified by the von Neumann entropy. Erasing memory involves transforming the initial state $\rho(0)$ into a low-entropy target state $\rho(\tau)$, leading to a decrease in system entropy $\Delta S=-\text{Tr}\left[\rho(\tau)\ln{\rho(\tau)}\right]+\text{Tr}\left[\rho(0)\ln{\rho(0)}\right]$. This entropy decrease is compensated by heat dissipated into the environment, which constitutes the thermodynamic cost of the erasure process $Q=-\int_{0}^{\tau}\text{Tr}\left[\dot{\rho}H(t)\right]{\rm d}t$. The second law of thermodynamics requires that the total entropy production $\Delta\Sigma=\Delta S+\beta Q$ be non-negative, which yields the general form of the Landauer principle $\beta Q\geq -\Delta S$. For the specific case of erasing one bit of information, this yields Eq.~\eqref{Landauer}.

To investigate the role of quantum coherence in memory erasure, we classify coherence into two categories based on the energy eigenbasis. First, the off-diagonal elements $\langle c_{j}|\rho|c'_{j'}\rangle$ with $c\neq c'$ represent coherence between different energy eigenspaces, which implies $[\rho,H]\neq 0$. Second, the off-diagonal elements with $c=c'$ and $j\neq j'$ represent coherence between degenerate states within the same energy eigenspace. A density matrix with only this type of coherence has a block-diagonal structure, which still commutes with $H$. The completely decohered state can be obtained by removing all off-diagonal elements, resulting in a strictly diagonal state $\rho_{\text{d}}=\sum_{c,j}\Pi_{c,j}\rho\Pi_{c,j}$, where $\Pi_{c,j}=|c_{j}\rangle\langle c_{j}|$ is the projection operator onto the individual energy eigenstate $|c_{j}\rangle$.


\textit{Finite-time Landauer bound.}---Information erasure is a process of state transformation, which can be characterized by the quantum speed limit between two states~\cite{Tan2021}. The corresponding quantum Landauer bound reads 
\begin{equation}\label{Landauer0}
  \beta Q\geq -\Delta S+\frac{\mathcal{W}_{q}[\rho(\tau),\rho(0)]^{2}}{\tau},
\end{equation}
where $\mathcal{W}_{q}$ denotes the quantum Wasserstein distance, which measures the minimal thermodynamic divergence between two states in a high-dimensional manifold of density matrices. In the regime where $\mathcal{W}_{q}^{2}/\tau\ll 1$, the second term on the right-hand side of inequality~\eqref{Landauer0} becomes negligible, and the bound approaches the Landauer bound. This condition can be conventionally achieved in the quasistatic limit. However, it can also be satisfied if the quantum Wasserstein distance $\mathcal{W}_{q}$ is intrinsically small even for a finite erasure time $\tau$.


Subsequently, we discuss how the different types of coherence affect the quantum Wasserstein distance and the thermodynamic cost. We first consider the coherence between different eigenspaces. The quantum Wasserstein distance is bounded from below by (see Supplemental Material (SM)~\cite{supp} for details)
\begin{equation}\label{Wasserstein1}
  \mathcal{W}_{q}[\rho(\tau),\rho(0)]^{2}\geq \frac{d[\rho(0),\rho(\tau)]^{2}}{\tau^{-1}\int_{0}^{\tau} \mathcal{G}[\rho(t)]{\rm d}t},
\end{equation}
where $\mathcal{G}(\rho):=(1/2)\sum_{k}\text{Tr}[\rho [L_{k},\phi]^{\dagger}[L_{k},\phi]]$. Here, we define $\phi:=\sum_{n}\text{sign}[p_{n}(\tau)-p_{n}(0)]|p_{n}(t)\rangle\langle p_{n}(t)|$, where $p_{n}(t)$ and $|p_{n}(t)\rangle$ are the eigenvalues and eigenstates of $\rho(t)=\sum_{n}p_{n}|p_{n}\rangle\langle p_{n}|$, respectively. Here, the trace-like distance is defined as $d[\rho(0),\rho(\tau)]:=\sum_{n}|p'_{n}(\tau)-p'_{n}(0)|$, where $\{p_{n}'(t)\}$ denotes the eigenvalues of $\rho(t)$ sorted in ascending order. According to the definition of $\phi$, we have $\phi^\dagger=\phi$, $\phi^2=\mathbb I$, and $[\rho,\phi]=0$. Using these relations and expanding the commutator~\cite{supp}, we obtain
\begin{equation}\label{Gcoherence}
  \mathcal{G}(\rho)=2\mathcal{A}(\rho)-\frac{1}{2}\sum_{k}\text{Tr}\left[\rho\{L_{k},\phi\}^{\dagger}\{L_{k},\phi\}\right],
\end{equation}
where $\mathcal{A}(\rho):=\sum_{k}\text{Tr}[\rho L_{k}^{\dagger}L_{k}]$ denotes the dynamical activity quantifying the timescale of thermal relaxation. The trace contribution in the second term on the right-hand side is non-negative. It can become nonzero when $[H(t),\rho(t)]\neq 0$ and the corresponding coherence connects eigenstates of $\rho$ whose populations change in opposite directions, indicating population backflow from target states to source states~\cite{supp}. Such coherence between different eigenspaces decreases $\mathcal{G}(\rho)$ and increases the lower bound of the quantum Wasserstein distance, resulting in an additional thermodynamic cost for memory erasure.

\begin{figure}[t]
  \centering
  \includegraphics[width=1\linewidth]{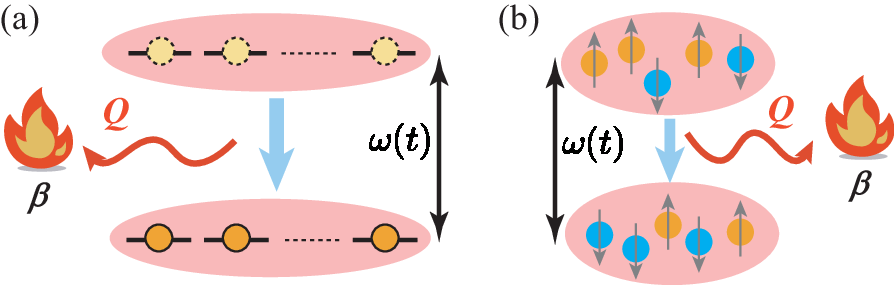}
  \caption{\label{Fig1} Schematics of finite-time memory erasure in (a) the $2N$-state model and (b) the collective spin model. The pink shaded regions indicate coherence among degenerate configurations within each effective energy level. In (a), each level contains $N$ degenerate states; in (b), the two logical states are the central Dicke states, with five spins shown for illustration. As the energy gap $\omega(t)$ is varied, transitions induced by the heat bath transfer population from the excited state to the ground state, dissipating heat $Q$ into the bath at inverse temperature $\beta$.}
  \end{figure}

We now examine the coherence between degenerate states. For the erasure processes considered here, in the absence of coherence between different eigenspaces, we have $\mathcal{G}(\rho)=2\mathcal{A}(\rho)$. We prove~\cite{supp} that the dynamical activity can be bounded from above by
\begin{equation}\label{activity1}
  \mathcal{A}(\rho)\leq \mathcal{A}(\rho_{\text{d}})+\mathcal{C}_{\mathcal{A}}\mathcal{C}_{l_{1}}(\rho),
\end{equation}
where $\mathcal{C}_{\mathcal{A}}:=\max_{c,j,j':j\neq j'}|\langle c_{j}|\sum_{k}L_{k}^{\dagger}L_{k}|c_{j'}\rangle|$ and $\mathcal{C}_{l_{1}}(\rho)$ denotes the widely used $l_{1}$-norm of coherence~\cite{Baumgratz2014, Streltsov2017},  defined as the sum of the absolute values of the off-diagonal elements of the density matrix in the energy eigenbasis: $\mathcal{C}_{l_{1}}(\rho):=\sum_{(c,j)\neq(c',j')}|\langle c_{j}|\rho|c'_{j'}\rangle|$. By combining Eqs.~\eqref{Landauer0}--\eqref{activity1}, we obtain the finite-time Landauer bound with only the coherence between degenerate states
\begin{equation}\label{Landauer1}
  \beta Q\geq -\Delta S+\frac{d[\rho(0),\rho(\tau)]^{2}}{2\displaystyle\int_{0}^{\tau}\left[\mathcal{A}(\rho_{\text{d}})+\mathcal{C}_{\mathcal{A}}\mathcal{C}_{l_{1}}(\rho)\right]{\rm d}t}.
\end{equation}
In the absence of coherence, only $\mathcal{A}(\rho_{\text{d}})$ survives, which suggests that the energy cost of the finite-time memory erasure cannot fall below the classical limit determined by $\mathcal{A}(\rho_{\text{d}})$. The presence of $\mathcal{C}_{\mathcal{A}}\mathcal{C}_{l_{1}}$ potentially increases dynamical activity, which allows these degenerate states to act as an entity and reduces the energy cost. Activating this mechanism requires two conditions: coherence between degenerate states must exist ($\mathcal{C}_{l_{1}}\neq 0$), and the jump operators $L_{k}$ must couple these degenerate states ($\mathcal{C}_{\mathcal{A}}\neq 0$). Provided this kind of coherence is sufficiently large, the penalty term in Eq.~\eqref{Landauer1} becomes negligible, potentially allowing the erasure process to approach the Landauer bound in finite time. In the following, we provide two examples, illustrated in Fig.~\ref{Fig1}, to demonstrate this mechanism.


\textit{$2N$-state model.}---Figure~\ref{Fig1}(a) shows the $2N$-state model, which comprises $N$ degenerate ground states $|g_j\rangle$ and $N$ degenerate excited states $|e_j\rangle$ ($j=1,2,\ldots,N$). Such an effective two-state model has been widely studied in thermometry~\cite{Correa2015,Scandi2025} and can be experimentally realized~\cite{Koch2007,HAFFNER2008,Saffman2010}. As illustrated in Fig.~\ref{Fig1}(a), information is erased within a finite time $\tau$ while the system is coupled to the heat bath and its energy gap is controlled by the time-dependent Hamiltonian
\begin{equation}\label{Hamiltonian}
  H(t)=\omega(t)\sum_{j=1}^{N}|e_{j}\rangle\langle e_{j}|,
\end{equation}
where $\omega(t)$ represents the time-dependent energy gap between the ground and excited states. The population transfer responsible for erasure is generated by the dissipative term in Eq.~\eqref{Lindblad}. The jump operators are given by $L_{0}=\sqrt{\kappa_{0}/N}\sum_{j,j'}|g_{j}\rangle\langle e_{j'}|$ and $L_{1}=\sqrt{\kappa_{1}/N}\sum_{j,j'}|e_{j}\rangle\langle g_{j'}|$~\cite{supp}. The coefficient $1/\sqrt{N}$ is similar to that used in the Dicke model to ensure a well-defined large-$N$ limit. This normalization keeps the total transition rate from each state independent of $N$. The transition rates $\kappa_{0}$ and $\kappa_{1}$ are determined by the heat bath and the coupling strengths between the system and the heat bath, which satisfy the detailed balance condition $\kappa_{1}=\kappa_{0}e^{-\beta\omega}$. In this setup, the jump operators act on degenerate states, satisfying the condition $\mathcal{C}_{\mathcal{A}}\neq 0$~\cite{supp}.

We first consider the initial state of the form $\rho(0)=p_{G}(0)|G\rangle\langle G|+p_{E}(0)|E\rangle\langle E|$ with $p_{G}(0)=p_{E}(0)=1/2$, where $|G\rangle:=\sum_{j}|g_{j}\rangle/\sqrt{N}$ and $|E\rangle:=\sum_{j}|e_{j}\rangle/\sqrt{N}$. This state is a steady state for $\omega(0)=0$. Such a state can be prepared experimentally in macroscopic collective systems~\cite{Abdurakhimov2016,Forn2019}. The density matrix retains a block-diagonal structure throughout the erasure process, with blocks containing coherence between the degenerate states. The corresponding von Neumann entropy of the initial state is $S[\rho(0)]=\ln{2}$, which encodes one bit of information. During the erasure process, the energy gap is varied according to a protocol $\omega(t)$ over the time interval $t\in[0,\tau]$, and the state of the system evolves following the Lindblad master equation~\eqref{Lindblad}. The target final state is characterized by $p_{E}(\tau)=\varepsilon$ and $p_{G}(\tau)=1-\varepsilon$, which is an almost zero-entropy state with an error probability $\varepsilon$ caused by the finite-time effect. For a given error probability $\varepsilon$, the amount of erased information is $-\Delta S=\ln{2}+\varepsilon\ln{\varepsilon}+(1-\varepsilon)\ln{(1-\varepsilon)}$, and the trace-like distance between the final state and the initial state is $d[\rho(0),\rho(\tau)]=1-2\varepsilon$. The initial coherence in $\rho(0)$ is treated as a pre-existing quantum resource that facilitates the erasure process~\cite{Streltsov2017}. Thus, we evaluate the thermodynamic cost of the erasure operation itself (erasing $\ln 2$ of information). This approach is justified as the dynamics preserves the block-diagonal structure, which ensures that the coherence between degenerate states is not consumed during the erasure process and remains available for subsequent computational cycles, including writing and erasing.

For the $2N$-state model, we show~\cite{supp} that the dynamical activity can be bounded from above by $\mathcal{A}\leq \kappa_{1}+\kappa_{0}+\kappa_{0}\mathcal{C}_{l_{1}}(\rho)$, with $\mathcal{C}_{l_{1}}(\rho)=N-1$ for the initial state $\rho(0)$, which remains constant over time. Therefore, the finite-time Landauer bound in Eq.~\eqref{Landauer1} evaluates to
\begin{equation}\label{Landauer_two_state1}
  \beta Q\geq -\Delta S+\frac{(1-2\varepsilon)^{2}}{2\displaystyle\int_{0}^{\tau}(\kappa_{1}+N\kappa_{0}){\rm d}t} =:\beta Q_{\text{b}},
\end{equation}
which suggests that the presence of coherence between degenerate states effectively enhances the transition from the excited states to the ground states. Equation~\eqref{Landauer_two_state1} further implies that the finite-time correction term vanishes as $O(N^{-1})$, which allows the thermodynamic cost to approach the Landauer bound in finite time for the large-$N$ limit.

\textit{Collective spin model.}---As another example, we consider an odd number $N_{s}$ of identical spins collectively coupled to a heat bath, illustrated in Fig.~\ref{Fig1}(b). With $J_{z}=\sum_{j=1}^{N_{s}}\sigma_{z}^{(j)}/2$, where $\sigma_{z}^{(j)}$ is the Pauli $z$ operator for the $j$th spin, the driven Hamiltonian is
\begin{equation}\label{Hamiltonian2}
  H(t)=\omega(t)J_{z}+\frac{U}{2}J_{z}^{2},
\end{equation}
which has been used to describe superabsorption~\cite{Kamimura2022}. The second term represents the interaction between the spins. The central transition has frequency $\omega(t)$, while the other transitions are detuned from it by $U$.

The symmetric Dicke state with $m$ excitations is $|D_{m}^{N_{s}}\rangle=\binom{N_{s}}{m}^{-1/2}\sum_{|x|=m}|x\rangle$, where the sum runs over all $N_s$-bit strings $x$ with Hamming weight $|x|=m$. As represented by the pink region in Fig.~\ref{Fig1}(b), the Dicke state contains the required coherence among degenerate spin configurations, which can be prepared experimentally~\cite{Lucke2014}. By coupling the system to a bath that suppresses the sideband transitions, the dynamics of the system is effectively confined to the two central Dicke states~\cite{Kamimura2022}. We choose $|G\rangle=|D_{k}^{N_{s}}\rangle$ and $|E\rangle=|D_{k+1}^{N_{s}}\rangle$ as the two logical states, where $k=(N_{s}-1)/2$. Their energy gap is $\omega(t)$, and the corresponding erasure process is similar to the description in the $2N$-state model. The state retains the form $\rho(t)=p_{G}(t)|G\rangle\langle G|+p_{E}(t)|E\rangle\langle E|$ throughout the erasure process, with $p_{G}(0)=p_{E}(0)=1/2$ and $p_{G}(\tau)=1-\varepsilon$, $p_{E}(\tau)=\varepsilon$. Hence, the erased information and the trace-like distance are the same as those in the $2N$-state model.

The collective jump operators are given by $L_{0}=\sqrt{\alpha\kappa_{0}}|G\rangle\langle E|$ and $L_{1}=\sqrt{\alpha\kappa_{1}}|E\rangle\langle G|$~\cite{supp}, where $\alpha=(N_{s}+1)^{2}/(4N_{s})$ is the collective enhancement factor. The factor $1/\sqrt{N_{s}}$ is introduced for the same reason as in the $2N$-state model. The transition rates $\kappa_{0}$ and $\kappa_{1}$ satisfy the detailed balance condition $\kappa_{1}=\kappa_{0}e^{-\beta\omega}$. In the local $\sigma_z$-product basis, the corresponding $l_{1}$-norm of coherence is $\mathcal{C}_{l_{1}}(\rho)=\binom{N_s}{k+1}-1$, which remains constant during the erasure process. The coherence is available for subsequent computation. Together with $\mathcal{C}_\mathcal{A}\neq0$~\cite{supp}, this satisfies the two conditions identified below Eq.~\eqref{Landauer1}. This bound therefore reveals the role of degenerate coherence, but it is not tight here because $\mathcal{C}_{l_1}$ counts all pairs of degenerate configurations, including those not connected by an allowed spin flip. Calculating the dynamical activity gives $\mathcal{A}(\rho)\leq\kappa_{1}+\alpha\kappa_0$~\cite{supp}. Using this result, the finite-time Landauer bound becomes
\begin{equation}\label{Landauer_two_state2}
\beta Q\geq-\Delta S+
\frac{(1-2\varepsilon)^2}
{2\displaystyle\int_0^\tau(\kappa_{1}+\alpha\kappa_0)\mathrm{d}t}.
\end{equation}
Since $\alpha=N_s/4+O(1)$, the finite-time correction vanishes as $O(N_s^{-1})$. Therefore, the lower bound on the thermodynamic cost approaches the Landauer bound in finite time as $N_s\to\infty$. This collective enhancement originates from the permutation symmetry of the Dicke states and collective coupling to the bath. Because the bath cannot distinguish which spin flips, the corresponding transition amplitudes interfere constructively.

\textit{The minimal energy cost.}---Inequalities~\eqref{Landauer_two_state1} and~\eqref{Landauer_two_state2} present the opportunity to approach the Landauer bound in finite-time memory erasure and elucidate its underlying physical mechanisms. A remaining question, however, is whether the bounds can be approached in finite-time erasure processes.


To investigate this, we study the minimal energy cost of the $2N$-state model by analyzing the heat functional for a given erasure time $\tau$ and target error probability $\varepsilon$. Considering the Hamiltonian~\eqref{Hamiltonian}, the master equation~\eqref{Lindblad} and the definition of heat, the heat functional with respect to $p_{E}$ can be given by $Q= \int_{\varepsilon}^{1/2}\omega\left(p_{E}\right){\rm d} p_{E}$. Focusing on the large-$N$ limit, the minimal cost is given by~\cite{Liu20253,supp}
\begin{equation}\label{Qmin}
  \beta Q_{\text{min}}(\tau,\varepsilon)=-\Delta S+ \left[ \int_{\varepsilon}^{\frac{1}{2}}\mathcal{J}(u)\mathrm{d}u \right]^{2}\frac{1}{N},
\end{equation}
where $\mathcal{J}(u)$ is determined by the transition rate $\kappa_{0}$ (see SM~\cite{supp} for its definition). This demonstrates that the minimal entropy production decreases as $N^{-1}$ in the large-$N$ limit, which suggests that the minimal thermodynamic cost approaches the Landauer bound in finite time as $N\rightarrow\infty$. Similar results can be obtained for the collective spin model~\cite{supp}. 






\begin{figure}[t]
  \centering
  \includegraphics[width=1\linewidth]{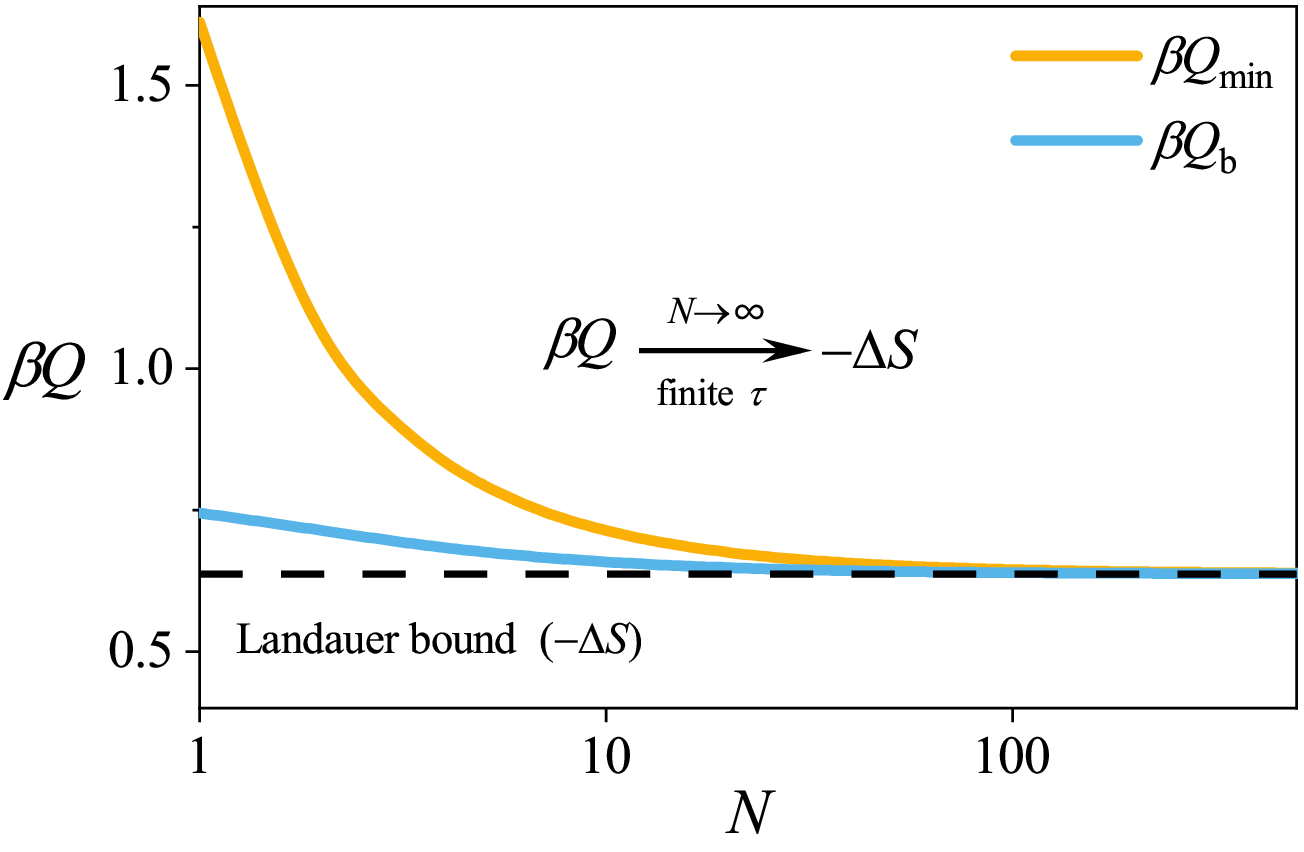}
  \caption{\label{Fig2} 
  Numerical illustration of the finite-time approach to the Landauer bound. The minimal thermodynamic cost $\beta Q_{\min}$ in Eq.~\eqref{Qmin} and the finite-time bound $\beta Q_{\rm b}$ defined in Eq.~\eqref{Landauer_two_state1} are plotted as functions of the degree of degeneracy $N$. The black dashed line represents the Landauer bound $-\Delta S$. As $N$ increases at a finite erasure time $\tau=1$, both quantities converge toward $-\Delta S$. Here $\beta=1$, $\mu=1$, and $\varepsilon=0.01$.}
\end{figure}

\textit{Continuous protocols.}---For continuous $\omega(t)$ on $[0,\tau]$, we prove for the $2N$-state model~\cite{supp} that
\begin{equation}\label{Q_continuous}
  \beta Q= -\Delta S+O(N^{-1}),
\end{equation}
in the large-$N$ limit. This result suggests that as $N\rightarrow\infty$, the continuous protocol can approach the Landauer bound in finite time with $N^{-1}$ behavior. For the collective spin model, we also obtain similar results~\cite{supp}. It is well known that a generic feature of optimal protocols is discontinuity at the initial moment~\cite{Tim2007, Salazar2019, Proesmans2020L, Liu20253}. Our framework recovers this jump behavior for finite $N$. Crucially, as $N$ increases, the magnitude of the initial discontinuity in the optimal protocol progressively diminishes and eventually vanishes as $N\rightarrow\infty$~\cite{supp}, which agrees with our findings for continuous protocols.


\textit{Illustration.}---To exemplify our results, we consider the $2N$-state model coupled to an Ohmic bosonic heat bath (see SM for the collective spin model). The transition rates of the Ohmic bosonic heat bath are given by $\kappa_{0}=\mu\omega(1+N_{\text{BE}})$ and $\kappa_{1}=\mu\omega N_{\text{BE}}$, where $\mu$ represents the coupling strength and $N_{\text{BE}}=1/(e^{\beta\omega}-1)$ denotes the Bose-Einstein distribution.

\begin{figure}[t]
  \centering
  \includegraphics[width=1\linewidth]{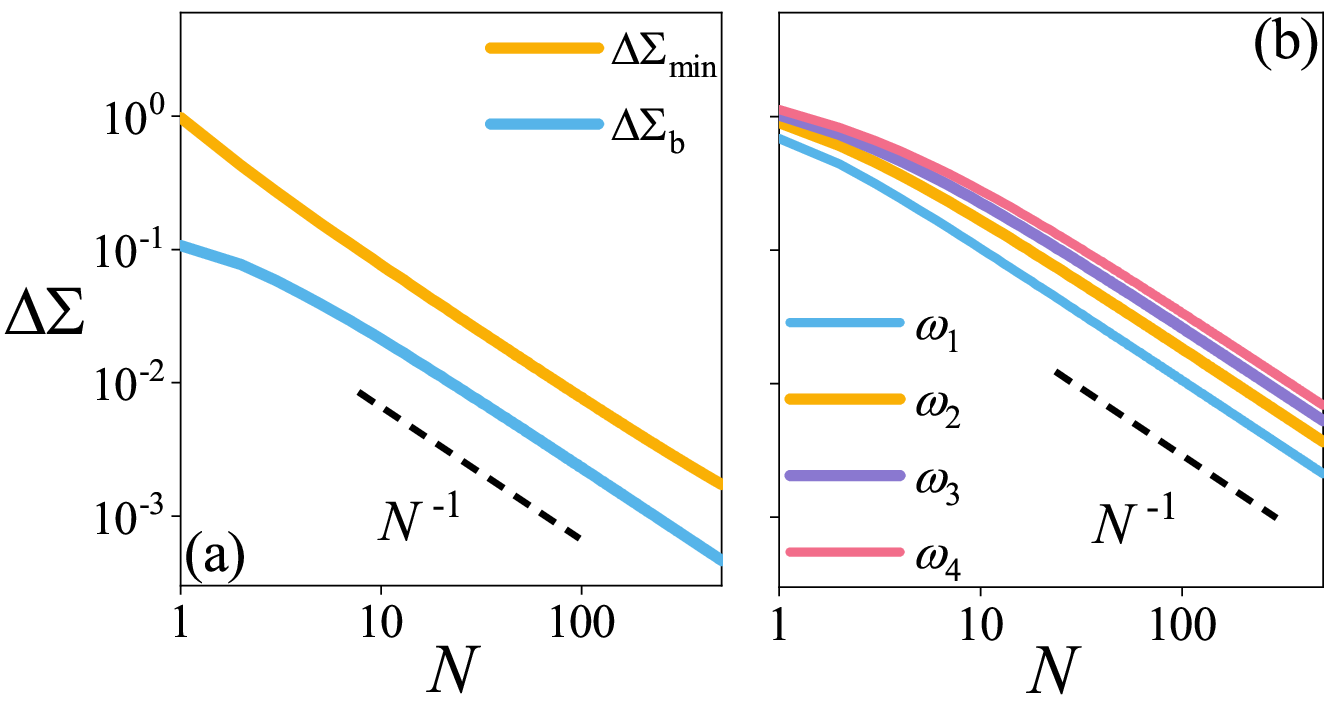}
  \caption{\label{Fig3} Entropy production as a function of the degree of degeneracy $N$. (a) Entropy production for the optimal protocol $\Delta \Sigma_{\text{min}}=\Delta S+\beta Q_{\text{min}}$ (yellow line) and $\Delta\Sigma_{\text{b}}=\Delta S+\beta Q_{\text{b}}$ (blue line). (b) Entropy production for four polynomial protocols $\omega_{n}(t)=5[1-(1-t/\tau)^{n}]$ ($n=1, 2, 3$ and $4$). The black dashed lines are drawn for $N^{-1}$ as the reference for both panels. Here $\beta=1$, $\mu=1$, $\tau=1$, and $\varepsilon=0.01$ in (a).
  }
\end{figure}

In Fig.~\ref{Fig2}, we present the minimal thermodynamic cost $\beta Q_{\min}$ and the finite-time bound $\beta Q_{\text{b}}$ as functions of $N$. The minimal thermodynamic cost consistently remains above the lower bound established by inequality~\eqref{Landauer_two_state1}. As $N$ increases at a fixed finite erasure time, both quantities converge toward the Landauer bound, demonstrating that the Landauer bound can be approached without increasing the operation time. This convergence persists for different values of $\tau$ and $\varepsilon$, as shown by additional numerical results in the SM~\cite{supp}. The corresponding penalty term is shown in Fig.~\ref{Fig3}(a). As predicted by inequality~\eqref{Landauer_two_state1} and Eq.~\eqref{Qmin}, the entropy production vanishes with $N^{-1}$ scaling in the large-$N$ limit.



To demonstrate Eq.~\eqref{Q_continuous}, we employ polynomial protocols $\omega_{n}(t)=5[1-(1-t/\tau)^{n}]$, which are continuous and differentiable over the interval $[0,\tau]$. The energy cost is determined by numerically solving the master equation~\eqref{Lindblad}. Figure~\ref{Fig3}(b) displays the resulting entropy production as a function of $N$ for four protocols ($n= 1, 2, 3$ and $4$). For all protocols tested, the entropy production clearly vanishes as $N^{-1}$ in the large-$N$ limit, verifying the robustness of our theoretical framework. The results for other parameters are provided in the SM~\cite{supp}, which also show the same behavior.

\textit{Summary.}---In summary, we show that coherence between degenerate states allows the Landauer bound to be approached at a finite erasure time, as demonstrated in both the $2N$-state model and the collective spin model. This finite-time advantage arises from resource substitution, with operation time replaced by scalable system size and coherence. This insight suggests potential pathways for designing energy-efficient electronic devices and is relevant to key tasks in quantum information processing and computation, such as the initialization of silicon electron-nuclear spin registers~\cite{Reiner2024} and the reset of superconducting qubits using quantum absorption refrigerators~\cite{Aamir2025,Blok2025}.

\begin{acknowledgments}
\textit{Acknowledgments.}---The authors thank the anonymous referees for their helpful comments. We are grateful to Tan Van Vu and Weitao Chen for useful discussion and critical reading of the manuscript. This work was financially supported by the National Natural Science Foundation of China (Grants No. 12475039, No. 12347151 and No. 12247172), and Guangdong Basic and Applied Basic Research Foundation (Grant No. 2025A1515010350). Y.L. gratefully acknowledges the Yukawa Research Fellow program, co-sponsored by the YITP and the Yukawa Memorial Foundation. 
\end{acknowledgments}

\textit{Data availability.}---The data are not publicly available upon publication. The data are available from the authors upon reasonable request.

\bibliography{References}

\end{document}


\bibliographystyle{apsrev4-2}

\title{Supplemental Material: \\Approaching the Landauer Bound without Slowing Down}

\author{Yue Liu}\thanks{These authors contributed equally to this work.}

\affiliation{Department of Physics and Shenzhen Research Institute, Xiamen University, Xiamen 361005, Fujian, China }

\affiliation{Yukawa Institute for Theoretical Physics, Kyoto University, Kitashirakawa Oiwakecho, Sakyo-ku, Kyoto 606-8502, Japan}

\author{Chenlong Huang}\thanks{These authors contributed equally to this work.}

\affiliation{Department of Physics and Shenzhen Research Institute, Xiamen University, Xiamen 361005, Fujian, China }

\author{Xingyu Zhang}

\affiliation{Department of Physics and Shenzhen Research Institute, Xiamen University, Xiamen 361005, Fujian, China }

\affiliation{School of Science, Zhejiang University of Science and Technology, Hangzhou 310023, China}

\author{Dahai He}\email{dhe@xmu.edu.cn}

\affiliation{Department of Physics and Shenzhen Research Institute, Xiamen University, Xiamen 361005, Fujian, China }

\date{\today}

\maketitle

\tableofcontents

\renewcommand{\theequation}{S\arabic{equation}}
\renewcommand{\thetable}{S\arabic{table}}
\renewcommand{\thefigure}{S\arabic{figure}}
\renewcommand{\thesection}{S\arabic{section}}

\section{Alternative expression of the Lindblad master equation}

The Lindblad master equation can be reformulated as
\begin{equation}\label{lindblad}
    \frac{d\rho}{dt}=\mathcal{K}_{\rho}\left(-\ln{\rho}+\ln{\rho^{\text{eq}}}\right),
\end{equation}
where $\mathcal{K}_{\rho}(x):=i\beta^{-1}[x,\rho]+\mathcal{O}_{\rho}(x)$. The superoperator $\mathcal{O}_{\rho}(x)$ is defined as
\begin{equation}
    \mathcal{O}_{\rho}(x):=\frac{1}{2}\sum_{k}e^{-\beta\omega_k/2}\left[L_{k},[\rho]_{\beta\omega_k}\left(\left[L_{k}^{\dagger},x\right]\right)\right],
\end{equation}
which corresponds to the Laplacian of a weighted graph. The tilted operator $[\rho]_{\theta}(X)$ is defined as a logarithmic mean transformation
\begin{equation}\label{tilted}
    [\rho]_{\theta}(X):=e^{-\theta/2}\int_{0}^{1}e^{s\theta}\rho^{s}X\rho^{1-s}ds=\sum_{n,m}\Phi\left(e^{\theta/2}p_{n},e^{-\theta/2}p_{m}\right)\left\langle p_n|X|p_m\right\rangle |p_n\left\rangle\right\langle p_m|,
\end{equation}
where $p_{n}$ and $|p_n\rangle$ are the eigenvalues and eigenstates of the density matrix $\rho$, i.e., $\rho=\sum_{n}p_{n}|p_n\rangle\langle p_n|$. The logarithmic mean of two positive numbers $x$ and $y$ is given by $\Phi(x,y)=\frac{x-y}{\ln{x}-\ln{y}}$ and $\Phi(x,x)=x$. As $\rho^{\text{eq}}=e^{-\beta H}/Z$ is the equilibrium state and $[\ln{\rho},\rho]=0$, one can immediately obtain
\begin{equation}
    i\beta^{-1}\left[-\ln{\rho}+\ln{\rho^{\text{eq}}},\rho\right]=-i[H,\rho],
\end{equation}
which is the first term of the Lindblad master equation. Hence, we shall focus on the second term
\begin{equation}
    \mathcal{O}_{\rho}(-\ln{\rho}+\ln{\rho^{\text{eq}}})=\sum_{k}\left[L_{k}\rho L_{k}^{\dagger}-\frac{1}{2}\left\{L_{k}^{\dagger}L_{k},\rho\right\}\right].
\end{equation}
This can be verified as follows
\begin{equation}
    \begin{split}
        \mathcal{O}_{\rho}(-\ln{\rho}+\ln{\rho^{\text{eq}}})&=\frac{1}{2}\sum_{k}e^{-\beta\omega_{k}/2}\left[L_{k},[\rho]_{\beta\omega_{k}}\left(\left[L_{k}^{\dagger},-\ln{\rho}+\ln{\rho^{\text{eq}}}\right]\right)\right]\\
        &=\frac{1}{2}\sum_{k}e^{-\beta\omega_{k}/2}\left[L_{k},[\rho]_{\beta\omega_{k}}\left(-\left[L_{k}^{\dagger},\ln{\rho}\right]-\beta\left[L_{k}^{\dagger},H\right]\right)\right]\\
        &\overset{(\text{a})}{=}\frac{1}{2}\sum_{k}e^{-\beta\omega_{k}/2}\left[L_{k},[\rho]_{\beta\omega_{k}}\left(\beta\omega_{k} L_{k}^{\dagger}-\left[L_{k}^{\dagger},\ln{\rho}\right]\right)\right]\\
        &\overset{(\text{b})}{=}\frac{1}{2}\sum_{k}e^{-\beta\omega_{k}/2}\left[L_{k},e^{\beta\omega_{k}/2}\rho L_{k}^{\dagger}-e^{-\beta\omega_{k}/2}L_{k}^{\dagger}\rho\right]\\
        &=\frac{1}{2}\sum_{k}\left[L_{k}\rho L_{k}^{\dagger}-\rho L_{k}^{\dagger}L_{k}+e^{-\beta\omega_{k}}\left(L_{k}^{\dagger}\rho L_{k}-L_{k}L_{k}^{\dagger}\rho\right)\right]\\
        &\overset{(\text{c})}{=}\frac{1}{2}\sum_{k}\left\{\left[L_{k}\rho L_{k}^{\dagger}-\rho L_{k}^{\dagger}L_{k}\right]+\left[L_{k}\rho L_{k}^{\dagger}-L_{k}^{\dagger}L_{k}\rho\right]\right\}\\
        &=\sum_{k}\left[L_{k}\rho L_{k}^{\dagger}-\frac{1}{2}\left\{L_{k}^{\dagger}L_{k},\rho\right\}\right],
    \end{split}
\end{equation}
which is the second term of the Lindblad master equation. In the above equations, (a) applies $\left[L_{k}^{\dagger},H\right]=-\omega_{k} L_{k}^{\dagger}$; (b) follows from the fact that $[\rho]_{\theta}([X,\ln{\rho}]-\theta X)=e^{-\theta/2}X\rho-e^{\theta/2}\rho X$ for arbitrary linear operators $X$ and $\theta\in\mathbb{R}$; (c) uses $L_{k}=e^{\beta\omega_{k}/2}L^{\dagger}_{k'}$, where the jump operator $L_{k'}$ describes the reverse jump process of $L_{k}$.

The inner product is defined as $\langle A,B\rangle=\text{Tr}[A^{\dagger}B]$ for arbitrary linear operators $A$ and $B$. One can prove that the inner product $\langle u,\mathcal{O}_{\rho}(v)\rangle$ satisfies linearity, conjugate-symmetry ($\langle u,\mathcal{O}_{\rho}(v)\rangle=\langle v,\mathcal{O}_{\rho}(u)\rangle^{*}$, where $*$ denotes the complex conjugate) and positive semi-definite properties ($\langle v,\mathcal{O}_{\rho}(v)\rangle\geq 0$). Therefore, $\langle u,\mathcal{O}_{\rho}(v)\rangle$ is also an inner product, which satisfies the Cauchy-Schwarz inequality
\begin{equation}\label{CSI}
    |\langle u,\mathcal{O}_{\rho}(v)\rangle|^{2}\leq \langle u,\mathcal{O}_{\rho}(u)\rangle\langle v,\mathcal{O}_{\rho}(v)\rangle.
\end{equation}

\section{Bounds of the quantum Wasserstein distance}
Let $\rho(t)=\sum_{n}p_{n}(t)|p_n(t)\rangle\langle p_n(t)|$ be a spectral decomposition of the density matrix, and define $\phi(t):=\sum_{n}\text{sgn}[p_{n}(\tau)-p_{n}(0)]|p_n(t)\rangle\langle p_n(t)|$, which is a self-adjoint operator and satisfies $\phi^{2}(t)=\mathbb{I}$ and $[\phi(t),\rho(t)]=0$. Here, $\text{sgn}(x)$ is the sign function, which is defined as $\text{sgn}(x)=1$ for $x\geq0$, $\text{sgn}(x)=-1$ otherwise. Hence, we have
\begin{equation}\label{sgn}
    \begin{split}
        \sum_{n}\text{sgn}[p_{n}(\tau)-p_{n}(0)][p_{n}(\tau)-p_{n}(0)]&=\text{Tr}\left[\int_{0}^{\tau}\phi(t)\dot{\rho}(t)dt\right]\\
        &\overset{(\text{a})}{=}\int_{0}^{\tau}\left\langle \phi,i\beta^{-1}[-\ln{\rho}+\ln{\rho^{\text{eq}}},\rho]+\mathcal{O}_{\rho}(-\ln{\rho}+\ln{\rho^{\text{eq}}})\right\rangle {\rm d}t\\
        &\overset{(\text{b})}{=}\int_{0}^{\tau}\left\langle \phi,\mathcal{O}_{\rho}(v)\right\rangle {\rm d}t ~~~~~~~~~~~~\left(v=-\ln{\rho}+\ln{\rho^{\text{eq}}}\right)\\
        &\overset{(\text{c})}{\leq}\left(\int_{0}^{\tau}\langle \phi,\mathcal{O}_{\rho}(\phi)\rangle {\rm d}t\right)^{1/2}\left(\int_{0}^{\tau}\langle v,\mathcal{O}_{\rho}(v)\rangle {\rm d}t\right)^{1/2}\\
        &\overset{(\text{d})}{\leq}\left(\tau^{-1}\int_{0}^{\tau}\langle \phi,\mathcal{O}_{\rho}(\phi)\rangle {\rm d}t\right)^{1/2}\left(\mathcal{W}_{q}[\rho(0),\rho(\tau)]^{2}+\delta\right)^{1/2}.
    \end{split}
\end{equation}
Here, (a) uses Eq.~\eqref{lindblad}; (b) follows from the fact $\langle\phi,[-\ln{\rho}+\ln{\rho^{\text{eq}}},\rho]\rangle=0$; (c) applies the Cauchy-Schwarz inequality~\eqref{CSI}; (d) uses the definition of the quantum Wasserstein distance
\begin{equation}
    \mathcal{W}_{q}[\rho(0),\rho(\tau)]^{2}+\delta\geq\tau\int_{0}^{\tau}\langle v,\mathcal{\mathcal{K}}_{\rho}(v)\rangle {\rm d}t,
\end{equation}
for a positive number $\delta$ and a smooth curve from $\rho(0)$ to $\rho(\tau)$. The inner product in the last line of Eq.~\eqref{sgn} can be rewritten by the definition of the inner product
\begin{equation}
    \begin{split}
    \langle \phi,\mathcal{O}_{\rho}(\phi)\rangle&=\frac{1}{2}\sum_{k}e^{-\beta\omega_{k}/2}\left\langle \phi,\left[L_{k},[\rho]_{\beta\omega_{k}}\left(\left[L_{k}^{\dagger},\phi\right]\right)\right]\right\rangle\\
    &=\frac{1}{2}\sum_{k}e^{-\beta\omega_{k}/2}\left\langle \left[L_{k}^{\dagger},\phi\right],[\rho]_{\beta\omega_{k}}\left(\left[L_{k}^{\dagger},\phi\right]\right)\right\rangle.
    \end{split} 
\end{equation}

Let $X_{k}:=\left[L_{k}^{\dagger},\phi\right]$, then we have
    \begin{equation}
        \begin{split}
            \langle \phi,\mathcal{O}_{\rho}(\phi)\rangle&\overset{(\text{a})}{=}\frac{1}{2}\sum_{k}e^{-\beta\omega_{k}/2}\sum_{n,m}\Phi\left(e^{\beta\omega_{k}/2}p_{n},e^{-\beta\omega_{k}/2}p_{m}\right)\langle n|X_{k}|m\rangle\langle m|X_{k}^{\dagger}|n\rangle\\
            &\overset{(\text{b})}{\leq}
            \frac{1}{4}\sum_{k}e^{-\beta\omega_{k}/2}\sum_{n,m}\left(e^{\beta\omega_{k}/2}p_{n}+e^{-\beta\omega_{k}/2}p_{m}\right)\langle n|X_{k}|m\rangle\langle m|X_{k}^{\dagger}|n\rangle\\
            &=\frac{1}{4}\sum_{k}e^{-\beta\omega_{k}/2}\left(\sum_{n,m} e^{\beta\omega_{k}/2}p_{n}\langle n|X_{k}|m\rangle\langle m|X_{k}^{\dagger}|n\rangle+\sum_{n,m} e^{-\beta\omega_{k}/2}p_{m}\langle m|X_{k}^{\dagger}|n\rangle \langle n|X_{k}|m\rangle\right)\\
            &\overset{(\text{c})}{=}\frac{1}{4}\sum_{k}e^{-\beta\omega_{k}/2}\left(\sum_{n} e^{\beta\omega_{k}/2}p_{n}\langle n|X_{k}X_{k}^{\dagger}|n\rangle+\sum_{m} e^{-\beta\omega_{k}/2}p_{m}\langle m|X_{k}^{\dagger}X_{k}|m\rangle\right)\\
            &=\frac{1}{4}\sum_{k}\left(\text{Tr}\{\rho X_{k}X_{k}^{\dagger}\}+e^{-\beta\omega_{k}}\text{Tr}\{\rho X_{k}^{\dagger}X_{k}\}\right)\\
            &=\frac{1}{2}\sum_{k}\text{Tr}\left\{\rho X_{k}X_{k}^{\dagger}\right\}:=\mathcal{G}(\rho).
        \end{split}
    \end{equation}
Here, (a) uses Eq.~\eqref{tilted}; (b) follows from the inequality of logarithmic mean $\Phi(x,y)\leq (x+y)/2$ for positive numbers $x$ and $y$; (c) applies the completeness relation $\sum_{n}|n\rangle\langle n|=\mathbb{I}$. 

Therefore, combining Eq.~\eqref{sgn} and the above inequality and letting $\delta\rightarrow 0$, one can obtain
\begin{equation}\label{bound}
   \mathcal{W}_{q}[\rho(\tau),\rho(0)]^{2}\geq \frac{d[\rho(0),\rho(\tau)]^{2}}{\tau^{-1}\int_{0}^{\tau} \mathcal{G}[\rho(t)]{\rm d}t},
\end{equation}
which is Eq.~(4) in the main text. Here, the trace-like distance is defined as $d[\rho(0),\rho(\tau)]:=\sum_{n}|p'_{n}(\tau)-p'_{n}(0)|$, where $\{p_{n}'(t)\}$ denotes the eigenvalues of $\rho(t)$ sorted in ascending order.

Direct expansion of the commutator and anticommutator,
together with \(\phi^\dagger=\phi\), \(\phi^2=\mathbb{I}\), and
\([\rho,\phi]=0\), yields
\begin{equation}
\begin{aligned}
\operatorname{Tr}\!\left\{
\rho[L_k^\dagger,\phi][L_k^\dagger,\phi]^\dagger
\right\}&=
2\operatorname{Tr}\!\left\{
\rho\left(
L_k^\dagger L_k+\phi L_k^\dagger L_k\phi
\right)\right\}
-\operatorname{Tr}\!\left\{
\rho\{L_k,\phi\}^\dagger\{L_k,\phi\}
\right\} \\
& =
4\operatorname{Tr}\!\left(
\rho L_k^\dagger L_k
\right)
-\operatorname{Tr}\!\left\{
\rho\{L_k,\phi\}^\dagger\{L_k,\phi\}
\right\}.
\end{aligned}
\end{equation}
The operator $\{L_{k}, \phi\}$ is positive semidefinite. Hence, we have
\begin{equation}
\mathcal{G}(\rho)= 2\mathcal{A}(\rho)-\frac{1}{2}\sum_{k}\text{Tr}[\rho \{L_{k},\phi\}^{\dagger}\{L_{k},\phi\}],
\end{equation}
where $\mathcal{A}(\rho)=\sum_{k}\text{Tr}[\rho L^{\dagger}_{k}L_{k}]$ is the dynamical activity. The second term is nonnegative since $\text{Tr}[AB]\geq0$ for positive semidefinite operators $A$ and $B$. A necessary condition for the term $\{L^{\dagger}_{k},\phi\}\{L_{k},\phi\}$ to vanish is $[L^{\dagger}_{k}L_{k},\phi]=0$. Note that $[L^{\dagger}_{k}L_{k}, H]=0$ holds due to $[L_{k},H]=\omega_{k} L_{k}$, implying that $L^{\dagger}_{k}L_{k}$ and $H$ share a common eigenbasis. If \(\phi\) mixes energy eigenspaces on which $L_k^\dagger L_k$ acts differently, then $[L_k^\dagger L_k,\phi]\neq0$, and the correction term in Eq. (S14) cannot vanish. The presence of quantum coherence between different energy eigenspaces implies that the density matrix does not commute with the Hamiltonian, i.e., $[H, \rho] \neq 0$, which leads to the existence of non-zero off-diagonal elements $H_{mn}:=\langle m| H|n\rangle\neq 0$. The matrix element of the commutator $[H,\phi]$ is given by
\begin{equation}
    \langle m | [H, \phi] | n\rangle = \langle m | (H\phi - \phi H) | n\rangle = H_{mn}(\phi_n - \phi_m),
\end{equation}
which means that if the coherence couples eigenstates associated with different population dynamics (i.e., $H_{mn} \neq 0$ while $\phi_m \neq \phi_n$), the commutator becomes non-zero. This shows that coherence between the subspaces of increasing $(\phi_n=1)$ and decreasing $(\phi_n=-1)$ populations can break the condition for the correction term to vanish, thereby increasing the lower bound of the quantum Wasserstein distance.


\section{Upper bound of dynamical activity}
H\"{o}lder's inequality reads
\begin{equation}
    |\langle A,B\rangle| \leq \|A\|_{p}\|B\|_{q},\quad \frac{1}{p}+\frac{1}{q}=1,
\end{equation}
where $\langle A,B\rangle=\text{Tr}[A^{\dagger}B]$ is the inner product of two matrices $A$ and $B$, and $\|\cdot\|_{p}$ is the entrywise $p$-norm of a matrix. This form of H\"{o}lder's inequality is equivalent to the vector form of H\"{o}lder's inequality by vectorizing the matrices.

Then we have
\begin{equation}
    \begin{split}
        \mathcal{A}(\rho)&=\sum_{k}\text{Tr}\left[\rho L^{\dagger}_{k}L_{k}\right]\\
        &=\text{Tr}\left[\rho_{\text{d}}\sum_{k}L^{\dagger}_{k}L_{k}\right]+\text{Tr}\left[(\rho-\rho_{\text{d}})\sum_{k}L^{\dagger}_{k}L_{k}\right] \\
        &\leq\text{Tr}\left[\rho_{\text{d}}\sum_{k}L^{\dagger}_{k}L_{k}\right]+\left|\text{Tr}\left[(\rho-\rho_{\text{d}})\sum_{k}L^{\dagger}_{k}L_{k}\right]\right| \\
        &\overset{(\text{a})}{\leq} \text{Tr}\left[\rho_{\text{d}}\sum_{k}L^{\dagger}_{k}L_{k}\right]+\mathcal{C}_\mathcal{A}\mathcal{C}_{l_{1}}(\rho)\\
        &=\mathcal{A}(\rho_{\text{d}})+\mathcal{C}_\mathcal{A}\mathcal{C}_{l_{1}}(\rho).
    \end{split}
\end{equation}
Here, (a) applies H\"{o}lder's inequality with $p=1$, $q=\infty$, $A=\rho-\rho_{\text{d}}$ and $B=\sum_{k}L^{\dagger}_{k}L_{k}$, where $\rho_{\text{d}}$ is the strictly diagonalized state defined in the main text. Here, $\mathcal{C}_{l_{1}}(\rho)$ denotes the $l_{1}$-norm of coherence of the density matrix $\rho(t)$ in the eigenbasis of the Hamiltonian $H$, defined as the sum of the absolute values of the off-diagonal elements, and $\mathcal{C}_\mathcal{A}$ is the infinity norm of the matrix $\sum_{k}L^{\dagger}_{k}L_{k}$.

Combining Eq.~\eqref{bound} and the above inequality, one can obtain
\begin{equation}
    \beta Q\geq -\Delta S+\frac{d[\rho(0),\rho(\tau)]^{2}}{2\displaystyle\int_{0}^{\tau}\left[\mathcal{A}(\rho_{\text{d}})+\mathcal{C}_{\mathcal{A}}\mathcal{C}_{l_{1}}(\rho)\right]{\rm d}t},
\end{equation}
which is Eq.~(7) in the main text.

\section{The $2N$-state model}
\subsection{Lindblad master equation}
The Hamiltonian of the total system reads
\begin{equation}
    H_{\text{tot}}=H+H_{B}+H_{I},
\end{equation}
where the system Hamiltonian is
\begin{equation}
    H=\omega\sum_{j=1}^{N}|e_{j}\rangle\langle e_{j}|
\end{equation}
with $|g_{j}\rangle$ and $|e_{j}\rangle$ being the $j$th ground and excited states of the system, respectively. The bath Hamiltonian depends on the specific physical model. For example, it can be modeled as a collection of harmonic oscillators
\begin{equation}
    H_{B}=\sum_{k}\Omega_{k}b_{k}^{\dagger}b_{k},
\end{equation}
where $b_{k}^{\dagger}$ and $b_{k}$ are the creation and annihilation operators of the $k$-th mode of the bath with frequency $\Omega_{k}$. The interaction Hamiltonian is given by
\begin{equation}
    H_{I}=\sum_{j, j', k}\frac{g}{\sqrt{N}}(\sigma_{j,j'}^{+}+\sigma_{j,j'}^{-})\otimes(b_{k}+b_{k}^{\dagger}),
\end{equation}
where $\sigma_{j,j'}^{+}=|e_{j}\rangle\langle g_{j'}|$ and $\sigma_{j,j'}^{-}=|g_{j'}\rangle\langle e_{j}|$ are the raising and lowering operators between the $j$-th excited state and the $j'$-th ground state, respectively, and $g$ is the coupling strength between the system and the bath. The coefficient $1/\sqrt{N}$ is introduced to ensure a fair comparison of the dynamics for different $N$.

After applying the Born-Markov and rotating-wave approximations, one can obtain the Lindblad master equation of the system, given by Eq.~(2) in the main text, where the jump operators are given by
\begin{equation}
    \begin{split}
    L_{0}=&\sqrt{\frac{\kappa_{0}}{N}}\sum_{j,j'}\sigma_{j,j'}^{-}=\sqrt{\frac{\kappa_{0}}{N}}\sum_{j,j'}|g_{j'}\rangle\langle e_{j}|,\\
    L_{1}=&\sqrt{\frac{\kappa_{1}}{N}}\sum_{j,j'}\sigma_{j,j'}^{+}=\sqrt{\frac{\kappa_{1}}{N}}\sum_{j,j'}|e_{j}\rangle\langle g_{j'}|.
    \end{split}
\end{equation}
This corresponds to $\mathcal{C}_{\mathcal{A}}=\kappa_{0}$.

\subsection{Memory erasure process}
We first consider the block-diagonal initial state
\begin{equation}
    \rho(0)=p_{G}|G\rangle\langle G|+p_{E}|E\rangle\langle E|,
\end{equation}
where $|G\rangle=\frac{1}{\sqrt{N}}\sum_{j=1}^{N}|g_{j}\rangle$ and $|E\rangle=\frac{1}{\sqrt{N}}\sum_{j=1}^{N}|e_{j}\rangle$. The density matrix is in a block-diagonal form in the basis of the Hamiltonian $H$ ($\{|e_{j}\rangle,|g_{j}\rangle;j=1,2,\cdots,N\}$) (see Eq.~\eqref{bdprocess}). The time evolution of $p_{G}$ and $p_{E}$ can be obtained from the Lindblad master equation
\begin{equation}\label{bdmaster}
    \begin{split}
        \frac{{\rm d}}{{\rm d}t}p_{G}&=-N\kappa_{1}p_{G}+N\kappa_{0}p_{E},\\
        \frac{{\rm d}}{{\rm d}t}p_{E}&=N\kappa_{1}p_{G}-N\kappa_{0}p_{E}.
    \end{split}
\end{equation}
As one can see, the transition rates are increased by a factor of $N$ compared to the two-state model, which can be regarded as a lubrication effect of the coherence among degenerate states. The initial state is $p_{G}=p_{E}=1/2$. We consider the erasure process that transfers the population from the excited subspace to the ground subspace in a finite time $\tau$, and erases the information encoded in the initial state, which is quantified by von Neumann entropy $S(\rho)=-\text{Tr}[\rho\ln{\rho}]$. The target final state is $p_{G}=1-\varepsilon$ and $p_{E}=\varepsilon$ with the error probability $\varepsilon\ll1$ due to the finite-time effects. The process can be described by
\begin{equation}\label{bdprocess}
    \renewcommand{\arraystretch}{1.2}
    \rho(0)=\frac{1}{N}
    \begin{pmatrix} 
        \frac{1}{2} & \frac{1}{2} & \cdots & \frac{1}{2} & 0 & 0 & \cdots & 0 \\ \frac{1}{2} & \frac{1}{2} & \cdots & \frac{1}{2} & 0 & 0 & \cdots & 0 \\ \vdots & \vdots & \ddots & \vdots & \vdots & \vdots & \ddots & \vdots \\ \frac{1}{2} & \frac{1}{2} & \cdots & \frac{1}{2} & 0 & 0 & \cdots & 0 \\ 0 & 0 & \cdots & 0 & \frac{1}{2} & \frac{1}{2} & \cdots & \frac{1}{2} \\ 0 & 0 & \cdots & 0 & \frac{1}{2} & \frac{1}{2} & \cdots & \frac{1}{2} \\ \vdots & \vdots & \ddots & \vdots & \vdots & \vdots & \ddots & \vdots \\ 0 & 0 & \cdots & 0 & \frac{1}{2} & \frac{1}{2} & \cdots & \frac{1}{2}
    \end{pmatrix}
    \xrightarrow{\text{erasure}}
    \rho(\tau)=\frac{1}{N}
    \begin{pmatrix} 
        \varepsilon & \varepsilon & \cdots & \varepsilon & 0 & 0 & \cdots & 0 \\ \varepsilon & \varepsilon & \cdots & \varepsilon & 0 & 0 & \cdots & 0 \\ \vdots & \vdots & \ddots & \vdots & \vdots & \vdots & \ddots & \vdots \\ \varepsilon & \varepsilon & \cdots & \varepsilon & 0 & 0 & \cdots & 0 \\ 0 & 0 & \cdots & 0 & 1-\varepsilon & 1-\varepsilon & \cdots & 1-\varepsilon \\ 0 & 0 & \cdots & 0 & 1-\varepsilon & 1-\varepsilon & \cdots & 1-\varepsilon \\ \vdots & \vdots & \ddots & \vdots & \vdots & \vdots & \ddots & \vdots \\ 0 & 0 & \cdots & 0 & 1-\varepsilon & 1-\varepsilon & \cdots & 1-\varepsilon
    \end{pmatrix},
\end{equation}
where the change of von Neumann entropy and the trace-like distance can be given by
\begin{equation}
    \Delta S=-[\ln{2}+\varepsilon\ln{\varepsilon}+(1-\varepsilon)\ln{(1-\varepsilon)}],~~~d[\rho(0),\rho(\tau)]=1-2\varepsilon.
\end{equation}
The $l_{1}$-norm of coherence during the erasure process is a constant which is given by
\begin{equation}
    \mathcal{C}_{l_{1}}(\rho)=\sum_{j\neq j'}\left|\langle g_{j}|\rho(t)|g_{j'}\rangle\right|+\sum_{j\neq j'}\left|\langle e_{j}|\rho(t)|e_{j'}\rangle\right|=N-1.
\end{equation}

\subsection{Upper bound of the dynamical activity}
The upper bound of the dynamical activity in the $2N$-state model reads
\begin{equation}
    \begin{split}
        \mathcal{A}(\rho)&\leq\mathcal{A}(\rho_{\text{d}})+\mathcal{C}_\mathcal{A}\mathcal{C}_{l_{1}}(\rho)\\
        &=\text{Tr}\left[L^{\dagger}_{0}L_{0}\rho_{\text{d}}\right]+\text{Tr}\left[L^{\dagger}_{1}L_{1}\rho_{\text{d}}\right]+\kappa_{0}\mathcal{C}_{l_{1}}(\rho)\\
        &=p_{G}\kappa_{1}+p_{E}\kappa_{0}+\kappa_{0}\mathcal{C}_{l_{1}}(\rho)\\
        &\overset{(a)}{\leq} \kappa_{0}+\kappa_{1}+\kappa_{0}\mathcal{C}_{l_{1}}(\rho),\\
    \end{split}
\end{equation}
where (a) uses the fact that $p_{G}\leq1$ and $p_{E}\leq1$. The lower bound of the thermodynamic cost of the erasure process can be given by
\begin{equation}
    \beta Q\geq -\Delta S+\frac{(1-2\varepsilon)^{2}}{2\displaystyle\int_{0}^{\tau}\left[\kappa_{0}+\kappa_{1}+\kappa_{0}\mathcal{C}_{l_{1}}(\rho)\right]{\rm d}t}.
\end{equation}
Therefore, the upper bound of the dynamical activity increases linearly with $N$ due to the coherence among degenerate states. Since $\mathcal{C}_{l_{1}}(\rho)=N-1$ for the information erasure process, the lower bound of the thermodynamic cost can be rewritten as
\begin{equation}
    \beta Q\geq -\Delta S+\frac{(1-2\varepsilon)^{2}}{2\displaystyle\int_{0}^{\tau}(\kappa_{1}+N\kappa_{0}){\rm d}t},
\end{equation}
which is Eq.~(9) in the main text. 


\section{The minimal energy cost}

\subsection{The general form}
Here, based on the results of our previous work, we consider the minimal energy cost of the erasure process in the $2N$-state model. The boundary conditions of the erasure process are $p_{E}(0)=1/2$ and $p_{E}(\tau)=\varepsilon$. According to the definition of heat dissipation $Q=-\int_{0}^{\tau}\text{Tr}[\dot{\rho}(t)H]{\rm d}t$ and the master equation Eq.~(2) (or Eq.~\eqref{bdmaster}), the heat dissipation can be expressed as a functional of the population of the excited states $p_{E}(t)$
\begin{equation}
    \beta Q=\int_{\varepsilon}^{1/2}\tilde{\omega}\left(p_{E},\frac{{\rm d}\tilde{t}}{{\rm d}p_{E}}\right) \mathrm{d} p_{E},
\end{equation}
where $\tilde{\omega}=\beta\omega$ and $\tilde{t}=t/\tau$ are the rescaled frequency and the rescaled time. By solving the Euler-Lagrange equation, one can determine the optimal $\omega(p_{E})$ that minimizes the heat dissipation by the following equation
\begin{equation}
        NCp_{E}R(\tilde{\omega})(1-e^{-\zeta})^{2}-\frac{R'(\tilde{\omega})}{R(\tilde{\omega})}(1-e^{-\zeta})- e^{-\zeta}=0,
\end{equation}
where $C$ is a constant determined by the boundary conditions and $\zeta=\tilde{\omega}-\ln{((1-p_{E})/p_{E})}$. Here, we define $\langle g_{j}|L_{0}L_{0}^{\dagger}|g_{j'}\rangle=\mu R(\tilde{\omega})$ for $\forall j,j'=1,2,\cdots,N$, where $R(\tilde{\omega})$ represents a rescaled transition rate and $\mu$ denotes the overall transition rate that is a constant in our study.

The inverse function $\tilde{t}(p_{E})$ of the optimal $p_{E}(\tilde{t})$ can be obtained by solving the above equation
\begin{equation}
    \tilde{t}(p_{E})=\frac{1}{\mu\tau}\int_{p_{E}}^{1/2}\frac{{\rm d}u}{uNR[\tilde{\omega}(u)](1-e^{-\zeta(u)})}.
\end{equation}
The above two equations are the parametric equations for $p_{E}$ ranging from $1/2$ to $\varepsilon$, which determine the optimal protocol $\tilde{\omega}(t)$ that minimizes the heat dissipation. The minimal heat dissipation can be calculated by substituting $\tilde{\omega}(p_{E})$ into the functional of heat dissipation
\begin{equation}
   \beta Q_{\min}=-\Delta S+\Delta\Sigma_{\min},
\end{equation}
where the second term is the minimal entropy production corresponding to the minimal heat dissipation. Once the jump operators are given, the optimal protocol and the minimal heat dissipation can be determined.

\subsection{The large-$N$ limit}

In the large-$N$ limit $N\gg1$, we have
    \begin{equation}
      \tilde{\omega}(p_{E}) \to \ln\frac{1 - p_{E}}{p_{E}},~\frac{\partial}{\partial \tilde{\omega}}\dot{p_{E}}(\tilde{\omega};p_{E}) = \frac{R^{\prime}(\tilde{\omega})}{R(\tilde{\omega})}\dot{p_{E}} - \mu N R(\tilde{\omega})e^{-\tilde{\omega}}(1 - p_{E}) \to - \mu p_{E} N R(\tilde{\omega}).
    \end{equation}
    We define
    \begin{equation}\label{eq:J}
      \mathcal{J}(x) = \left[x\mu\tau R\left( \ln\dfrac{1 - x}{x} \right) \right]^{-\frac{1}{2}}.
    \end{equation}
    Then, one can obtain
    \begin{equation}
      \sqrt{\frac{N\mu\tau}{C}} \to \int_{\varepsilon}^{\frac{1}{2}}\mathcal{J}(u)\mathrm{d}u,
    \end{equation}
    and
    \begin{equation}
      \tilde{t}(p_{E}) \to \displaystyle\left[ \int_{\varepsilon}^{\frac{1}{2}}\mathcal{J}(u)\mathrm{d}u \right]^{-1}\int_{p_{E}}^{\frac{1}{2}}\mathcal{J}(u)\mathrm{d}u.
    \end{equation}

    In the large-$N$ limit, one can obtain
    \begin{equation}
      \sqrt{\frac{NC}{\mu\tau}}\zeta \to \mathcal{J}(p_{E}),
    \end{equation}
    which implies
    \begin{equation}
      \zeta(p_{E}) = \sqrt{\frac{NC}{\mu\tau}}\zeta \cdot  \sqrt{\frac{N\mu\tau}{C}} \cdot \frac{1}{N} \sim \left[ J(p_{E})\int_{\varepsilon}^{\frac{1}{2}}\mathcal{J}(u)\mathrm{d}u \right]\frac{1}{N}.
    \end{equation}
    The optimal protocol can be written as
    \begin{equation}
      \begin{cases}
        \tilde{\omega}(p_{E}) = \ln\dfrac{1 - p_{E}}{p_{E}} + \left[ \mathcal{J}(p_{E})\displaystyle\int_{\varepsilon}^{\frac{1}{2}}\mathcal{J}(u)\mathrm{d}u \right]\dfrac{1}{N},\\
        \\
        \tilde{t}(p_{E}) = \displaystyle\left[ \int_{\varepsilon}^{\frac{1}{2}}\mathcal{J}(u)\mathrm{d}u \right]^{-1}\int_{p_{E}}^{\frac{1}{2}}\mathcal{J}(u)\mathrm{d}u,
      \end{cases}
    \end{equation}
    and the minimal heat production reads
    \begin{equation}\label{minimal}
      \begin{split}
        \beta Q_{\min} &= \int_{\varepsilon}^{\frac{1}{2}}\tilde{\omega}(p_{E})\mathrm{d}p_{E}  = -\Delta S + \left[ \int_{\varepsilon}^{\frac{1}{2}}\mathcal{J}(u)\mathrm{d}u \right]^{2}\frac{1}{N},
      \end{split}
    \end{equation}
    where
    \begin{equation}
      \Delta S = -\int_{\varepsilon}^{\frac{1}{2}}\ln\frac{1 - p_{E}}{p_{E}}\mathrm{d}p_{E} = -\ln2 - \varepsilon\ln\varepsilon - (1 - \varepsilon)\ln(1 - \varepsilon),
    \end{equation}
    which is the entropy change during the erasure process. Equation~\eqref{minimal} corresponds to Eq. (12) in the main text and shows that the excess cost above the Landauer bound scales as $O(N^{-1})$ in the large-$N$ limit.

\section{The continuous protocols}

The non-negative protocol $\tilde{\omega}$ is right-continuous and smooth almost everywhere, i.e., allowing for a finite number of discontinuities. Let $t_{i}$ be the $i$-th discontinuity point, and $\tilde{\omega}_{i}^{-} = \tilde{\omega}(t_{i}^{-})$ and $\tilde{\omega}_{i}^{+} = \tilde{\omega}(t_{i}^{+}) = \tilde{\omega}(t_{i})$ be the left limit and right limit of the protocol at the $i$-th discontinuity point $t_{i}$, respectively.
The system master equation is
\begin{equation}
    \dot{p}_{E} = - \tilde{\Gamma} p_{E} + \tilde{\Gamma} e^{-\tilde{\omega}} (1 - p_{E}),
\end{equation}
where $\tilde{\Gamma}$ satisfies $\tilde{\Gamma} \to +\infty$ as $N \to \infty$. Let $\tilde{\Gamma}(N,\tilde{\omega}) = \Gamma(N)R(\tilde{\omega})$.
The solution to the equation is
\begin{equation}
    p_{E}(t) = p_{E}(0) e^{- \Gamma \int_{0}^{t} R(\tilde{\omega}(s))(1 + e^{-\tilde{\omega}(s)}) \mathrm{d}s} + \Gamma \int_{0}^{t} R(\tilde{\omega}(r))e^{-\tilde{\omega}(r)} e^{- \Gamma \int_{r}^{t} R(\tilde{\omega}(s))(1 + e^{-\tilde{\omega}(s)}) \mathrm{d}s} \mathrm{d}r,
\end{equation}
where
\begin{equation}
    0 \le p_{E}(0) e^{- \Gamma \int_{0}^{t} R(\tilde{\omega}(s))(1 + e^{-\tilde{\omega}(s)}) \mathrm{d}s} \le p_{E}(0) e^{- \Gamma \int_{0}^{t} R(\tilde{\omega}(s)) \mathrm{d}s},
\end{equation}
that is, the first term is $O(e^{-\Gamma \int_{0}^{t} R(\tilde{\omega}(s)) \mathrm{d}s})$, and the main contribution is from the second term,
\begin{equation}
\begin{split}
    \Gamma \int_{0}^{t} R(\tilde{\omega}(r))e^{-\tilde{\omega}(r)} e^{- \Gamma \int_{r}^{t} R(\tilde{\omega}(s))(1 + e^{-\tilde{\omega}(s)}) \mathrm{d}s} \mathrm{d}r = \Gamma \int_{0}^{t} R(\tilde{\omega}(t - r))e^{-\tilde{\omega}(t - r)} e^{\Gamma \int_{t}^{t - r} R(\tilde{\omega}(s))(1 + e^{-\tilde{\omega}(s)}) \mathrm{d}s} \mathrm{d}r.
\end{split}
\end{equation}
Let $\phi(x) = R(\tilde{\omega}(t - x))e^{-\tilde{\omega}(t - x)}$ and $f(x) = \int_{t}^{t - x} R(\tilde{\omega}(s))(1 + e^{-\tilde{\omega}(s)}) \mathrm{d}s$. Then the second term is
\begin{equation}
\Gamma \int_{0}^{t} \phi(r) e^{\Gamma f(r)} \mathrm{d}r.
\end{equation}
Obviously, $f(x)$ reaches its absolute maximum at $x = 0$; thus
\begin{equation}
\Gamma \int_{0}^{t} \phi(r) e^{\Gamma f(r)} \mathrm{d}r = \Gamma \int_{0}^{\delta} \phi(r) e^{\Gamma f(r)} \mathrm{d}r + \Gamma \int_{\delta}^{t} \phi(r) e^{\Gamma f(r)} \mathrm{d}r,
\end{equation}
where
\begin{equation}
0 \le \Gamma \int_{\delta}^{t} \phi(r) e^{\Gamma f(r)} \mathrm{d}r \le \Gamma e^{\Gamma f(\delta)}\int_{\delta}^{t} R(\tilde{\omega}(s)) \mathrm{d}s.
\end{equation}
Here, $\delta$ is chosen such that $\lim_{N \to \infty} \delta = 0$ and $\lim_{N \to \infty} \Gamma f(\delta) = -\infty$. We can choose $\delta = \Gamma^{-\frac{1}{2}}$. Therefore,
\begin{equation}
p_{E}(t) = \Gamma \int_{0}^{\delta} \phi(r) e^{\Gamma f(r)} \mathrm{d}r + O(e^{-\Gamma \int_{0}^{t} R(\tilde{\omega}(s)) \mathrm{d}s}) + O(\Gamma e^{\Gamma f(\delta)}).
\end{equation}
Since $y = f(x)$ is monotonic and analytic, its inverse function $x = \varphi(y)$ exists. By Lagrange inversion, we can obtain
\begin{equation}
    \begin{split}
\varphi(y) &= \lim_{\zeta \to 0} \sum_{n = 1}^{\infty} \dfrac{1}{n!} \dfrac{\mathrm{d}^{n - 1}}{\mathrm{d}\zeta^{n - 1}} \left\{ \left[ \dfrac{\zeta}{f(\zeta) - f(0)} \right]^{n} \right\} y^{n}\\
&= - \dfrac{1}{R(\tilde{\omega}(t))(1 + e^{-\tilde{\omega}(t)})} y + \dfrac{[R^{\prime}(\tilde{\omega}(t))(1 + e^{-\tilde{\omega}(t)}) - R(\tilde{\omega}(t))e^{-\tilde{\omega}(t)}]\tilde{\omega}^{\prime}(t)}{2R^{3}(\tilde{\omega}(t))(1 + e^{-\tilde{\omega}(t)})^{3}} y^{2} + O(y^{3}),
\end{split}
\end{equation}
thus
\begin{equation}
    \begin{split}
    \phi(\varphi(y)) &= \phi(0) + \lim_{\zeta \to 0} \sum_{n = 1}^{\infty} \dfrac{1}{n!} \dfrac{\mathrm{d}^{n - 1}}{\mathrm{d}\zeta^{n - 1}} \left\{ \phi^{\prime}(\zeta) \left[ \dfrac{\zeta}{f(\zeta) - f(0)} \right]^{n} \right\} y^{n}\\
    &= R(\tilde{\omega}(t))e^{-\tilde{\omega}(t)} + \dfrac{[R^{\prime}(\tilde{\omega}(t)) - R(\tilde{\omega}(t))]e^{-\tilde{\omega}(t)}\tilde{\omega}^{\prime}(t)}{R(\tilde{\omega}(t))(1 + e^{-\tilde{\omega}(t)})} y + O(y^{2}).
    \end{split}
\end{equation}
Therefore, we have
\begin{equation}
    \begin{split}
    p_{E}(t) &= \Gamma \int_{0}^{f(\delta)} \phi(\varphi(y)) \varphi^{\prime}(y) e^{\Gamma y} \mathrm{d}y + O(e^{-\Gamma \int_{0}^{t} R(\tilde{\omega}(s)) \mathrm{d}s}) + O(\Gamma e^{\Gamma f(\delta)})\\
    &= \Gamma \int_{0}^{f(\delta)} \left[ -\dfrac{e^{-\tilde{\omega}(t)}}{1 + e^{-\tilde{\omega}(t)}} + \dfrac{\tilde{\omega}^{\prime}(t)e^{-\tilde{\omega}(t)}}{R(\tilde{\omega}(t)) (1 + e^{-\tilde{\omega}(t)})^{3}} y + O(y^{2}) \right] e^{\Gamma y} \mathrm{d}y + O(e^{-\Gamma \int_{0}^{t} R(\tilde{\omega}(s)) \mathrm{d}s}) + O(\Gamma e^{\Gamma f(\delta)})\\
    &= \int_{0}^{\Gamma f(\delta)} \left[ -\dfrac{e^{-\tilde{\omega}(t)}}{1 + e^{-\tilde{\omega}(t)}} + \dfrac{\tilde{\omega}^{\prime}(t)e^{-\tilde{\omega}(t)}}{R(\tilde{\omega}(t)) (1 + e^{-\tilde{\omega}(t)})^{3}} \dfrac{u}{\Gamma} + O\left( \dfrac{u^{2}}{\Gamma^{2}} \right) \right] e^{u} \mathrm{d}u + O(e^{-\Gamma \int_{0}^{t} R(\tilde{\omega}(s)) \mathrm{d}s}) + O(\Gamma e^{\Gamma f(\delta)})\\
    &= \dfrac{e^{-\tilde{\omega}(t)}}{1 + e^{-\tilde{\omega}(t)}} + \dfrac{\tilde{\omega}^{\prime}(t)e^{-\tilde{\omega}(t)}}{R(\tilde{\omega}(t)) (1 + e^{-\tilde{\omega}(t)})^{3}} \dfrac{1}{\Gamma} + O(e^{\Gamma f(\delta)}) + O\left( \dfrac{1}{\Gamma^{2}} \right) + O(e^{-\Gamma \int_{0}^{t} R(\tilde{\omega}(s)) \mathrm{d}s}) + O(\Gamma e^{\Gamma f(\delta)})\\
    &= \dfrac{e^{-\tilde{\omega}(t)}}{1 + e^{-\tilde{\omega}(t)}} + \dfrac{\tilde{\omega}^{\prime}(t)e^{-\tilde{\omega}(t)}}{R(\tilde{\omega}(t)) (1 + e^{-\tilde{\omega}(t)})^{3}} \dfrac{1}{\Gamma} + O\left( \dfrac{1}{\Gamma^{2}} \right).
    \end{split}
\end{equation}
Thus, in the limit $\Gamma \to +\infty$ (or $N \to \infty$), we have
\begin{align}
    \beta Q =& \int_{p_{E}(\tau)}^{p_{E}(0)} \tilde{\omega} \mathrm{d}p_{E}\notag\\
    =& \int_{p_{E}(t_{1}^{-})}^{p_{E}(0)} \tilde{\omega} \mathrm{d}p_{E} + \sum_{i = 1}^{M - 1} \int_{p_{E}(t_{i + 1}^{-})}^{p_{E}(t_{i}^{+})} \tilde{\omega} \mathrm{d}p_{E} + \int_{p_{E}(\tau)}^{p_{E}(t_{M}^{+})} \tilde{\omega} \mathrm{d}p_{E}\notag\\
    =&~\tilde{\omega}(0) p_{E}(0) - \tilde{\omega}(t_{1}^{-}) p_{E}(t_{1}^{-}) + \sum_{i = 1}^{M - 1} \tilde{\omega}(t_{i}^{+}) p_{E}(t_{i}^{+}) - \tilde{\omega}(t_{i + 1}^{-}) p_{E}(t_{i + 1}^{-}) + \tilde{\omega}(t_{M}^{+}) p_{E}(t_{M}^{+})\notag\\
    &- \tilde{\omega}(\tau) p_{E}(\tau) + \int_{\tilde{\omega}(0)}^{\tilde{\omega}(t_{1}^{-})} p_{E} \mathrm{d}\tilde{\omega} + \sum_{i = 1}^{M - 1} \int_{\tilde{\omega}(t_{i}^{+})}^{\tilde{\omega}(t_{i + 1}^{-})} p_{E} \mathrm{d}\tilde{\omega} + \int_{\tilde{\omega}(t_{M}^{+})}^{\tilde{\omega}(\tau)} p_{E} \mathrm{d}\tilde{\omega}\notag\\
    =&~\tilde{\omega}(0) p_{E}(0) + \sum_{i = 1}^{M} (\tilde{\omega}_{i}^{+} - \tilde{\omega}_{i}^{-}) p_{E}(t_{i}^{-}) - \tilde{\omega}(\tau) p_{E}(\tau) + \left[ \int_{\tilde{\omega}(0)}^{\tilde{\omega}_{1}^{-}} + \sum_{i = 1}^{M - 1} \int_{\tilde{\omega}_{i}^{+}}^{\tilde{\omega}_{i + 1}^{-}} + \int_{\tilde{\omega}_{M}^{+}}^{\tilde{\omega}(\tau)} \right] p_{E} \mathrm{d}\tilde{\omega}\notag\\
    =&~\tilde{\omega}(0) p_{E}(0) - \tilde{\omega}(\tau) p_{E}(\tau) + \sum_{i = 1}^{M} (\tilde{\omega}_{i}^{+} - \tilde{\omega}_{i}^{-}) p_{E}(t_{i}^{-})\notag\\
    &+ \left[ \int_{\tilde{\omega}(0)}^{\tilde{\omega}_{1}^{-}} + \sum_{i = 1}^{M - 1} \int_{\tilde{\omega}_{i}^{+}}^{\tilde{\omega}_{i + 1}^{-}} + \int_{\tilde{\omega}_{M}^{+}}^{\tilde{\omega}(\tau)} \right] \dfrac{e^{-\tilde{\omega}}}{1 + e^{-\tilde{\omega}}} + \dfrac{e^{-\tilde{\omega}}}{t^{\prime}(\tilde{\omega}) R(\tilde{\omega}) (1 + e^{-\tilde{\omega}})^{3}} \dfrac{1}{\Gamma} \mathrm{d}\tilde{\omega} + O\left( \dfrac{1}{\Gamma^{2}} \right)\notag\\
    =&~\tilde{\omega}(0) p_{E}(0) - \tilde{\omega}(\tau) p_{E}(\tau) - \ln\dfrac{1 + e^{-\tilde{\omega}(\tau)}}{1 + e^{-\tilde{\omega}(0)}} + \sum_{i = 1}^{M} \left[(\tilde{\omega}_{i}^{+} - \tilde{\omega}_{i}^{-}) p_{E}(t_{i}^{-}) - \ln\dfrac{1 + e^{-\tilde{\omega}_{i}^{-}}}{1 + e^{-\tilde{\omega}_{i}^{+}}}\right]\notag\\
    &+ \left[ \int_{\tilde{\omega}(0)}^{\tilde{\omega}_{1}^{-}} + \sum_{i = 1}^{M - 1} \int_{\tilde{\omega}_{i}^{+}}^{\tilde{\omega}_{i + 1}^{-}} + \int_{\tilde{\omega}_{M}^{+}}^{\tilde{\omega}(\tau)} \right] \dfrac{e^{-\tilde{\omega}}}{t^{\prime}(\tilde{\omega}) R(\tilde{\omega}) (1 + e^{-\tilde{\omega}})^{3}} \dfrac{1}{\Gamma} \mathrm{d}\tilde{\omega} + O\left( \dfrac{1}{\Gamma^{2}} \right)\notag\\
    =& -\Delta S + \boxed{\sum_{i = 1}^{M} \left[ (\tilde{\omega}_{i}^{+} - \tilde{\omega}_{i}^{-}) \dfrac{e^{-\tilde{\omega}_{i}^{-}}}{1 + e^{-\tilde{\omega}_{i}^{-}}} - \ln\dfrac{1 + e^{-\tilde{\omega}_{i}^{-}}}{1 + e^{-\tilde{\omega}_{i}^{+}}} \right]}\notag\\
    &+ \dfrac{1}{\Gamma} \left\{ \boxed{\sum_{i = 1}^{M} \dfrac{(\tilde{\omega}_{i}^{+} - \tilde{\omega}_{i}^{-}) e^{-\tilde{\omega}_{i}^{-}}}{t^{\prime}(\tilde{\omega}_{i}^{-}) R(\tilde{\omega}_{i}^{-}) (1 + e^{-\tilde{\omega}_{i}^{-}})^{3}}} + \boxed{\left[ \int_{\tilde{\omega}(0)}^{\tilde{\omega}_{1}^{-}} + \sum_{i = 1}^{M - 1} \int_{\tilde{\omega}_{i}^{+}}^{\tilde{\omega}_{i + 1}^{-}} + \int_{\tilde{\omega}_{M}^{+}}^{\tilde{\omega}(\tau)} \right] \dfrac{e^{-\tilde{\omega}}}{t^{\prime}(\tilde{\omega}) R(\tilde{\omega}) (1 + e^{-\tilde{\omega}})^{3}} \mathrm{d}\tilde{\omega}} \right\} + O\left( \dfrac{1}{\Gamma^{2}} \right).
\end{align}
Here, $M$ denotes the total count of discontinuous points. For continuous protocol, noting that $\Gamma(N)=N$ in our case, we have
\begin{equation}
    \beta Q=-\Delta S + O(N^{-1}),
\end{equation}
which is Eq.~(13) in the main text.

\section{The collective spin model}
\label{sec:collective_spin_model}
\subsection{Lindblad master equation}
To show that the mechanism discussed in the main text is not specific to the $2N$-state model, we consider an odd number $N_s$ of identical spins collectively coupled to a heat bath. The Hamiltonian of the total system reads
\begin{equation}\label{eq:spin_total_hamiltonian}
    H_{\rm tot}=H+H_B+H_I,
\end{equation}
where the driven system Hamiltonian is
\begin{equation}
    H=\omega J_z+\frac{U}{2}J_z^2,
    \qquad
    J_z=\frac{1}{2}\sum_{j=1}^{N_s}\sigma_z^{(j)}.
    \label{eq:spin_system_hamiltonian}
\end{equation}
The bath is modeled as a collection of bosonic modes,
\begin{equation}
    H_{B}=\sum_{k}\Omega_{k}b_{k}^{\dagger}b_{k},
    \label{eq:spin_bath_hamiltonian}
\end{equation}
where $b_k^\dagger$ and $b_k$ are the creation and annihilation operators of the $k$th bath mode with frequency $\Omega_k$. We take the interaction Hamiltonian to be
\begin{equation}
    H_I=\frac{1}{\sqrt{N_s}}(J_++J_-)
    \otimes\sum_q\left(g_q b_q+g_q^*b_q^\dagger\right),
    \qquad
    J_\pm=\sum_{j=1}^{N_s}\sigma_\pm^{(j)},
    \label{eq:spin_interaction_hamiltonian}
\end{equation}
where the factor $1/\sqrt{N_s}$ plays the same role as the factor $1/\sqrt{N}$ in
the $2N$-state model.

We restrict the dynamics to the fully symmetric subspace with total spin $J=N_s/2$, which is invariant under both $H(t)$ and $J_\pm$. A convenient basis of this subspace is formed by the symmetric Dicke states
\begin{equation}
    \begin{split}
    |D_m^{N_s}\rangle
        &=\binom{N_s}{m}^{-1/2}\sum_{|x|=m}|x\rangle,\\
    J_z|D_m^{N_s}\rangle
        &=\left(m-\frac{N_s}{2}\right)|D_m^{N_s}\rangle,\\
    J_-|D_{m+1}^{N_s}\rangle
        &=\sqrt{(m+1)(N_s-m)}\,|D_m^{N_s}\rangle,
    \end{split}
    \label{eq:dicke_states_and_matrix_element}
\end{equation}
where $m=0,1,\ldots,N_s$ and the sum runs over all $N_s$-bit strings $x$ with Hamming weight $|x|=m$. These states diagonalize $H(t)$. Their energies and the signed gaps between neighboring states are
\begin{equation}
    \begin{split}
    E_m(t)&=\omega(t)\left(m-\frac{N_s}{2}\right)
    +\frac{U}{2}\left(m-\frac{N_s}{2}\right)^2,\\
    \Delta_m(t)&=E_{m+1}(t)-E_m(t)
    =\omega(t)+U(m-k),
    \qquad k=\frac{N_s-1}{2}.
    \end{split}
    \label{eq:dicke_energies}
\end{equation}
Thus, the central transition
$|D_{k+1}^{N_s}\rangle\leftrightarrow|D_k^{N_s}\rangle$ has the energy gap
$\Delta_k(t)=\omega(t)$, while the other transitions form sidebands displaced
by integer multiples of $U$ in the signed spectrum. The physical frequency of
a sideband is $|\Delta_m(t)|$.

As in the superabsorption setup, we consider a frequency-selective bath. Its
filter response $F(\nu)$ is assumed to retain the central transition and
suppress all sideband transitions throughout the protocol,
\begin{equation}
    F\!\left[\omega(t)\right]\simeq1,
    \qquad
    F\!\left[|\Delta_m(t)|\right]\simeq0
    \quad (m\neq k).
    \label{eq:spin_filter_condition}
\end{equation}
This condition requires the central band and the sidebands to remain
spectrally resolved for all $t$. We assume that $U$ and the bath passband are
chosen such that this condition holds over the entire range swept by
$\omega(t)$.

Within the weak-coupling Markovian regime, we assume that the drive changes negligibly over the bath correlation time. Applying the Born--Markov and rotating-wave approximations at each time then gives the Lindblad master equation in Eq.~(2) of the main text. After the sidebands are neglected, only transitions between the two central excitation sectors remain resonant. We denote the projector onto
the $m$-excitation sector by
\begin{equation}
    \Pi_m=\sum_{|x|=m}|x\rangle\langle x|.
    \label{eq:spin_excitation_projector}
\end{equation}
The jump operators acting on the full central sectors are
\begin{equation}
    L_0(t)=\sqrt{\frac{\kappa_0(t)}{N_s}}\,
    \Pi_kJ_-\Pi_{k+1},
    \qquad
    L_1(t)=\sqrt{\frac{\kappa_1(t)}{N_s}}\,
    \Pi_{k+1}J_+\Pi_k.
    \label{eq:spin_jump_operators}
\end{equation}
Here, $\kappa_0(t)$ and $\kappa_1(t)$ are the downward and upward rates of the central transition, satisfying the detailed-balance condition $\kappa_1(t)=e^{-\beta\omega(t)}\kappa_0(t)$.

Within the fully symmetric subspace, only the two central Dicke states
\begin{equation}
    |G\rangle=|D_k^{N_s}\rangle,
    \qquad
    |E\rangle=|D_{k+1}^{N_s}\rangle
    \label{eq:central_dicke_states}
\end{equation}
remain coupled. According to 
\begin{equation}
    J_-|E\rangle
    =\sqrt{(k+1)(N_s-k)}\,|G\rangle
    =\frac{N_s+1}{2}\,|G\rangle,\qquad
    J_+|G\rangle
    =\sqrt{(k+1)(N_s-k)}\,|E\rangle
    =\frac{N_s+1}{2}\,|E\rangle,
    \label{eq:spin_jump_matrix_element}
\end{equation}
the collective matrix elements of the jump operators are enhanced by a factor
\begin{equation}
    \frac{1}{\sqrt{N_s}}\langle G|J_-|E\rangle
    =\frac{N_s+1}{2\sqrt{N_s}}
    =\sqrt{\alpha},
    \qquad
    \frac{1}{\sqrt{N_s}}\langle E|J_+|G\rangle
    =\frac{N_s+1}{2\sqrt{N_s}}
    =\sqrt{\alpha},
    \label{eq:spin_collective_factor}
\end{equation}
where $\alpha=(N_s+1)^2/(4N_s)$, which approaches $N_s/4$ for large $N_s$. Accordingly, the restrictions of Eq.~\eqref{eq:spin_jump_operators} to the symmetric subspace are
\begin{equation}
    L_0(t)=\sqrt{\alpha\kappa_0(t)}\,
    |G\rangle\langle E|,
    \qquad
    L_1(t)=\sqrt{\alpha\kappa_1(t)}\,
    |E\rangle\langle G|.
    \label{eq:spin_effective_jump_operators}
\end{equation}

\subsection{Memory erasure process}
We consider the initial state
\begin{equation}
    \rho(0)=p_G(0)|G\rangle\langle G|
    +p_E(0)|E\rangle\langle E|.
    \label{eq:spin_block_initial_state}
\end{equation}
In the local $\sigma_z$-product basis, which diagonalizes $H(t)$, the density matrix is block diagonal (see Eq.~\eqref{eq:spin_erasure_process}), while each block contains coherence between the degenerate product states. Since Eq.~\eqref{eq:spin_jump_operators} maps $|E\rangle$ to $|G\rangle$ and vice versa, the state retains the form $\rho(t)=p_G(t)|G\rangle\langle G|+p_E(t)|E\rangle\langle E|$ throughout the erasure process. The populations obey
\begin{equation}
    \begin{split}
        \frac{{\rm d}}{{\rm d}t}p_G
        &=-\alpha\kappa_1p_G
        +\alpha\kappa_0p_E,\\
        \frac{{\rm d}}{{\rm d}t}p_E
        &=\alpha\kappa_1p_G
        -\alpha\kappa_0p_E.
    \end{split}
    \label{eq:spin_population_dynamics}
\end{equation}
Thus, the transition rates are enhanced by the collective factor $\alpha$, which scales linearly with $N_s$.

We set $p_G(0)=p_E(0)=1/2$ and consider an erasure process that transfers the population from $|E\rangle$ to $|G\rangle$ within a finite time $\tau$. The target state satisfies $p_G(\tau)=1-\varepsilon$ and $p_E(\tau)=\varepsilon$, where $\varepsilon\ll1$ is the final error probability. Let $d_m=\binom{N_s}{m}$ and let $\boldsymbol{0}_{d_m}$ denote the $d_m\times d_m$ zero matrix. Since $N_s$ is odd, the two central excitation sectors have the same dimension $d=d_k=d_{k+1}$. Defining $\boldsymbol{u}_d=d^{-1/2}(1,1,\ldots,1)^{\mathsf T}$ and $E_d=\boldsymbol{u}_d\boldsymbol{u}_d^\dagger$, the erasure process $\rho(0)\xrightarrow{\mathrm{erasure}}\rho(\tau)$ can be written in the full local product basis, ordered by Hamming weight, as
\begin{equation}
    \begin{aligned}
    \rho(0)={}&
    \left(\bigoplus_{m=0}^{k-1}\boldsymbol{0}_{d_m}\right)
    \oplus\frac{1}{2}E_d\oplus\frac{1}{2}E_d
    \oplus\left(\bigoplus_{m=k+2}^{N_s}\boldsymbol{0}_{d_m}\right),\\
    \rho(\tau)={}&
    \left(\bigoplus_{m=0}^{k-1}\boldsymbol{0}_{d_m}\right)
    \oplus(1-\varepsilon)E_d\oplus\varepsilon E_d
    \oplus\left(\bigoplus_{m=k+2}^{N_s}\boldsymbol{0}_{d_m}\right).
    \end{aligned}
    \label{eq:spin_erasure_process}
\end{equation}
The change of von Neumann entropy and the trace-like distance are
\begin{equation}
    \Delta S=-\left[\ln2+\varepsilon\ln\varepsilon
    +(1-\varepsilon)\ln(1-\varepsilon)\right],
    \qquad
    d[\rho(0),\rho(\tau)]=1-2\varepsilon.
    \label{eq:spin_entropy_and_distance}
\end{equation}

At any intermediate time, the two central blocks have weights $p_G(t)$ and $p_E(t)$, while all other blocks remain zero. It follows that the $l_1$-norm of coherence remains constant during erasure:
\begin{equation}
    \begin{split}
    \mathcal{C}_{l_1}(\rho)
    &=p_G(t)(d-1)+p_E(t)(d-1)\\
    &=d-1.
    \end{split}
    \label{eq:spin_l1_coherence}
\end{equation}

\subsection{Upper bound of the dynamical activity}


For $\omega(t)\geq0$, detailed balance gives $\kappa_1\leq\kappa_0$. In the local product basis, the largest off-diagonal matrix element of $L_0^\dagger L_0+L_1^\dagger L_1$ is
\begin{equation}
    \mathcal{C}_{\mathcal A}=\frac{\kappa_0}{N_s}.
    \label{eq:spin_activity_coefficient}
\end{equation}
The upper bound on the dynamical activity based on the $l_1$-norm of coherence consequently becomes
\begin{equation}
    \mathcal{A}(\rho)
    \leq\frac{N_s+1}{2N_s}\left(\kappa_0p_E+\kappa_1p_G\right)
    +\frac{\kappa_0}{N_s}(d-1).
    \label{eq:spin_l1_activity_bound}
\end{equation}
This bound shows explicitly that both $\mathcal{C}_{l_1}\neq0$ and $\mathcal{C}_{\mathcal A}\neq0$ are present. It is not tight for the Dicke states because $\mathcal{C}_{l_1}$ counts all pairs of degenerate product states, whereas $J_\pm$ directly connects only configurations that differ by one spin flip.

A tighter bound follows by evaluating the activity of the symmetric Dicke
state itself:
\begin{equation}
    \begin{split}
    \mathcal{A}(\rho)
    &=\alpha\left(\kappa_0p_E+\kappa_1p_G\right)\\
    &\leq\kappa_{1}+\alpha\kappa_0.
    \end{split}
    \label{eq:spin_exact_activity}
\end{equation}
Substituting Eq.~\eqref{eq:spin_exact_activity} into the finite-time Landauer
bound gives
\begin{equation}
    \beta Q\geq-\Delta S+
    \frac{(1-2\varepsilon)^2}
    {2\displaystyle\int_0^\tau(\kappa_{1}+\alpha\kappa_0){\rm d}t}.
    \label{eq:spin_finite_time_landauer}
\end{equation}
Since
$\alpha=N_s/4+1/2+1/(4N_s)$, the finite-time correction vanishes as $O(N_s^{-1})$.

\subsection{The optimal and continuous protocols}
\label{sec:spin_protocols}

The optimal-protocol calculation and continuous-protocol analysis for the collective-spin model follow the same procedure as those for the $2N$-state model. Indeed, Eq.~\eqref{eq:spin_population_dynamics} has the same form as the population dynamics of the $2N$-state model under the replacement
$N\rightarrow\alpha$. In the large-$N_s$ limit,
\begin{equation}
    \alpha
    =\frac{(N_s+1)^2}{4N_s}
    =\frac{N_s}{4}+O(1).
    \label{eq:spin_alpha_large}
\end{equation}
Writing $\tilde{\omega}=\beta\omega$ and
$\kappa_0(\tilde{\omega})=\mu R(\tilde{\omega})$, we define
\begin{equation}
    \mathcal{J}(x)=
    \left[
    x\mu\tau R\!\left(\ln\frac{1-x}{x}\right)
    \right]^{-1/2},
    \label{eq:spin_J_function}
\end{equation}
which is the same as Eq.~\eqref{eq:J} for the $2N$-state model. For the boundary conditions $p_E(0)=1/2$ and $p_E(\tau)=\varepsilon$, the minimum heat dissipation in the large-$N_s$
limit is
\begin{equation}
    \begin{split}
    \beta Q_{\min}
    &=-\Delta S+
    \frac{1}{\alpha}
    \left[
    \int_{\varepsilon}^{1/2}\mathcal{J}(u)\,{\rm d}u
    \right]^2
    +o\!\left(\alpha^{-1}\right)\\
    &=-\Delta S+
    \frac{4}{N_s}
    \left[
    \int_{\varepsilon}^{1/2}\mathcal{J}(u)\,{\rm d}u
    \right]^2
    +o\!\left(N_s^{-1}\right).
    \end{split}
    \label{eq:spin_Qmin}
\end{equation}
Similarly, for a continuous protocol $\omega(t)$ on $[0,\tau]$ satisfying the same boundary conditions, the heat dissipation obeys
\begin{equation}
    \beta Q
    =-\Delta S+O\!\left(\alpha^{-1}\right)
    =-\Delta S+O\!\left(N_s^{-1}\right).
    \label{eq:spin_Qcontinuous}
\end{equation}
Thus, both the optimal protocol and continuous protocols approach the Landauer bound with $N_s^{-1}$ scaling.

\section{Numerical verification}
In this section, we present more numerical results to verify the analytical results of the $2N$-state model and the collective spin model. 
\subsection{The $2N$-state model}
\begin{figure}[htbp]
\centering
\includegraphics[width=0.8\textwidth]{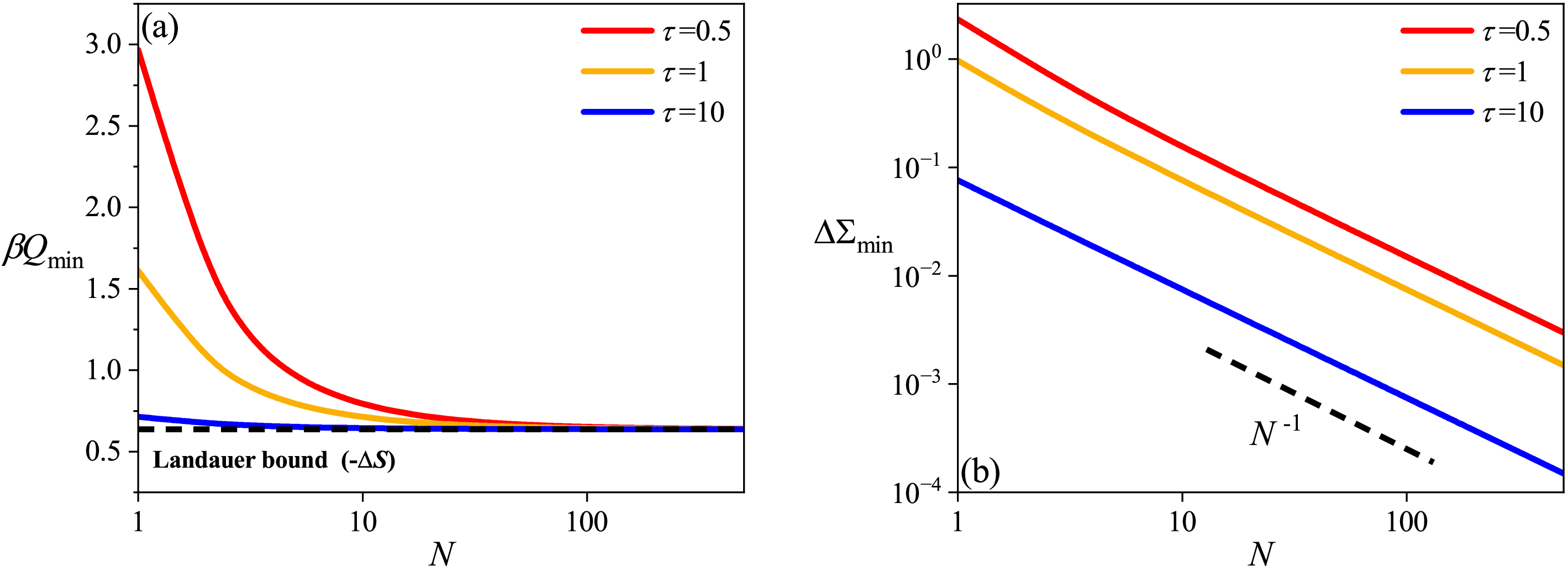}
\caption{Numerical results of the $2N$-state model. (a) The minimal heat dissipation $\beta Q_{\text{min}}$ as a function of $N$ for different erasure times $\tau$. (b) The minimal entropy production $\Delta\Sigma_{\text{min}}$ as a function of  $N$ for different erasure times $\tau$. The black dashed line in (a) represents the Landauer bound $-\Delta S$, and the black dashed line in (b) guides the eye for the scaling of $N^{-1}$. Here, $\beta=1$, $\mu=1$ and $\varepsilon=0.01$.}
\label{FigS1}
\end{figure}

Here, we present the numerical results of the $2N$-state model. The numerical results are obtained by solving the master equation Eq.~(2) in the main text. In Fig.~\ref{FigS1}(a), we show the minimal heat dissipation $\beta Q_{\text{min}}$ as a function of the degree of degeneracy $N$ for different erasure times $\tau$. We consider three erasure times, $\tau=0.5$, $1$, and $10$. The numerical results show that the minimal heat dissipation $\beta Q_{\text{min}}$ approaches the Landauer bound $-\Delta S$ as $N$ increases for different erasure times $\tau$. Although the minimal heat dissipation $\beta Q_{\text{min}}$ for the short erasure time $\tau=0.5$ for $N=1$ is very far from the Landauer bound $-\Delta S$, it also approaches the Landauer bound $-\Delta S$ as $N$ increases. In Fig.~\ref{FigS1}(b), we show the minimal entropy production $\Delta\Sigma_{\text{min}}=\beta Q_{\text{min}}+\Delta S$ as a function of the degree of degeneracy $N$ for different erasure times $\tau$ that we use in Fig.~\ref{FigS1}(a). The numerical results show that the minimal entropy production $\Delta\Sigma_{\text{min}}$ approaches zero as $N^{-1}$ for different erasure times $\tau$. The numerical results of the $2N$-state model verify the analytical results in the main text.

\begin{figure}[htbp]
\centering
\includegraphics[width=0.8\textwidth]{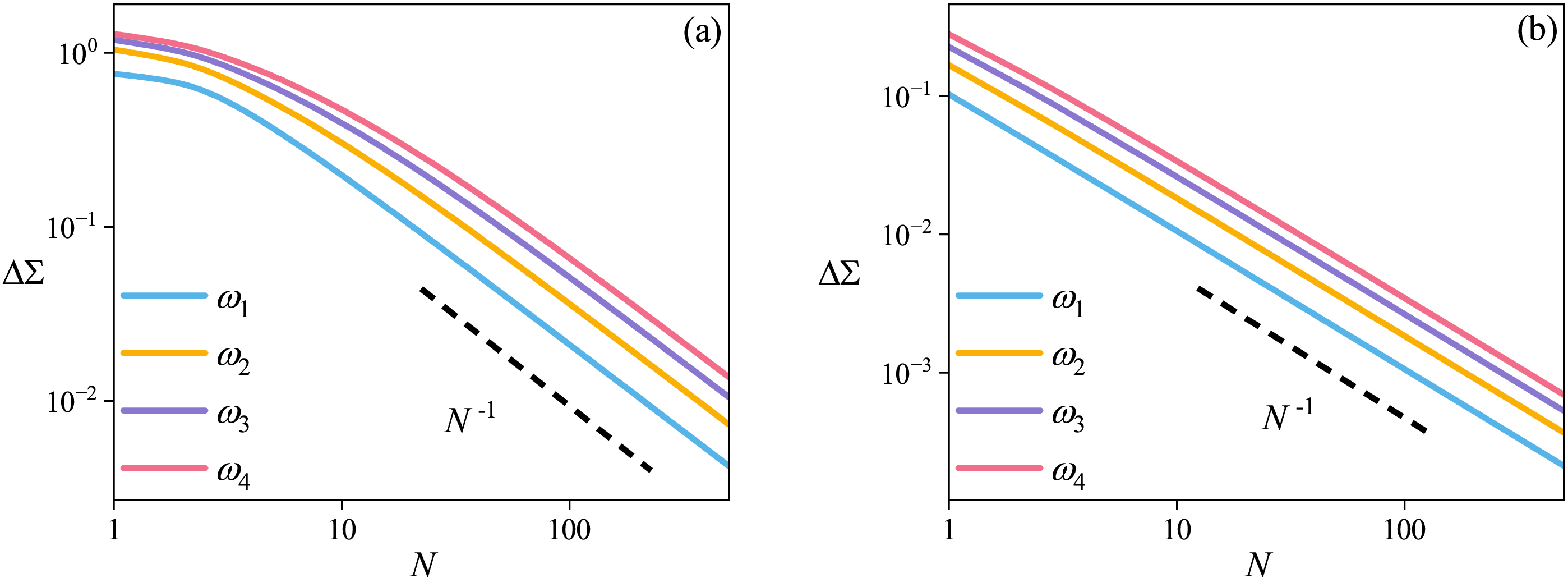}
\caption{Numerical results of the polynomial protocol for the $2N$-state model. The entropy production $\Delta\Sigma$ as a function of $N$ for four polynomial protocols with different values of $n$ ($n=1,2,3,4$) with the same erasure time $\tau=0.1$ in (a) and $\tau=10$ in (b). The black dashed lines guide the eye for the scaling of $N^{-1}$. Here, $\beta=1$ and $\mu=1$.}
\label{FigS2}
\end{figure}

Then, we present the result of the polynomial protocol 
\begin{equation}
    \omega_{n}(t)=5\left[1-\left(1-\frac{t}{\tau}\right)^{n}\right]
\end{equation}
for the $2N$-state model. In Fig.~\ref{FigS2}(a), we show the entropy production $\Delta\Sigma$ as a function of $N$ for four polynomial protocols with different values of $n$ ($n=1,2,3,4$) with the same erasure time $\tau=0.1$. The numerical results show that the entropy production $\Delta\Sigma$ approaches zero as $N^{-1}$ for polynomial protocols with different values of $n$. In Fig.~\ref{FigS2}(b), we also show the entropy production $\Delta\Sigma$ as a function of $N$ for four different polynomial protocols for $n$ ($n=1,2,3,4$) with the erasure time $\tau=10$, which also shows that the entropy production $\Delta\Sigma$ approaches zero as $N^{-1}$ for polynomial protocols with different values of $n$. The numerical results of the polynomial protocol verify the analytical results in the main text. Of course, one can also choose other continuous protocols, such as the exponential protocol, to verify the analytical results in the main text.

\begin{figure}[htbp]
\centering
\includegraphics[width=0.4\textwidth]{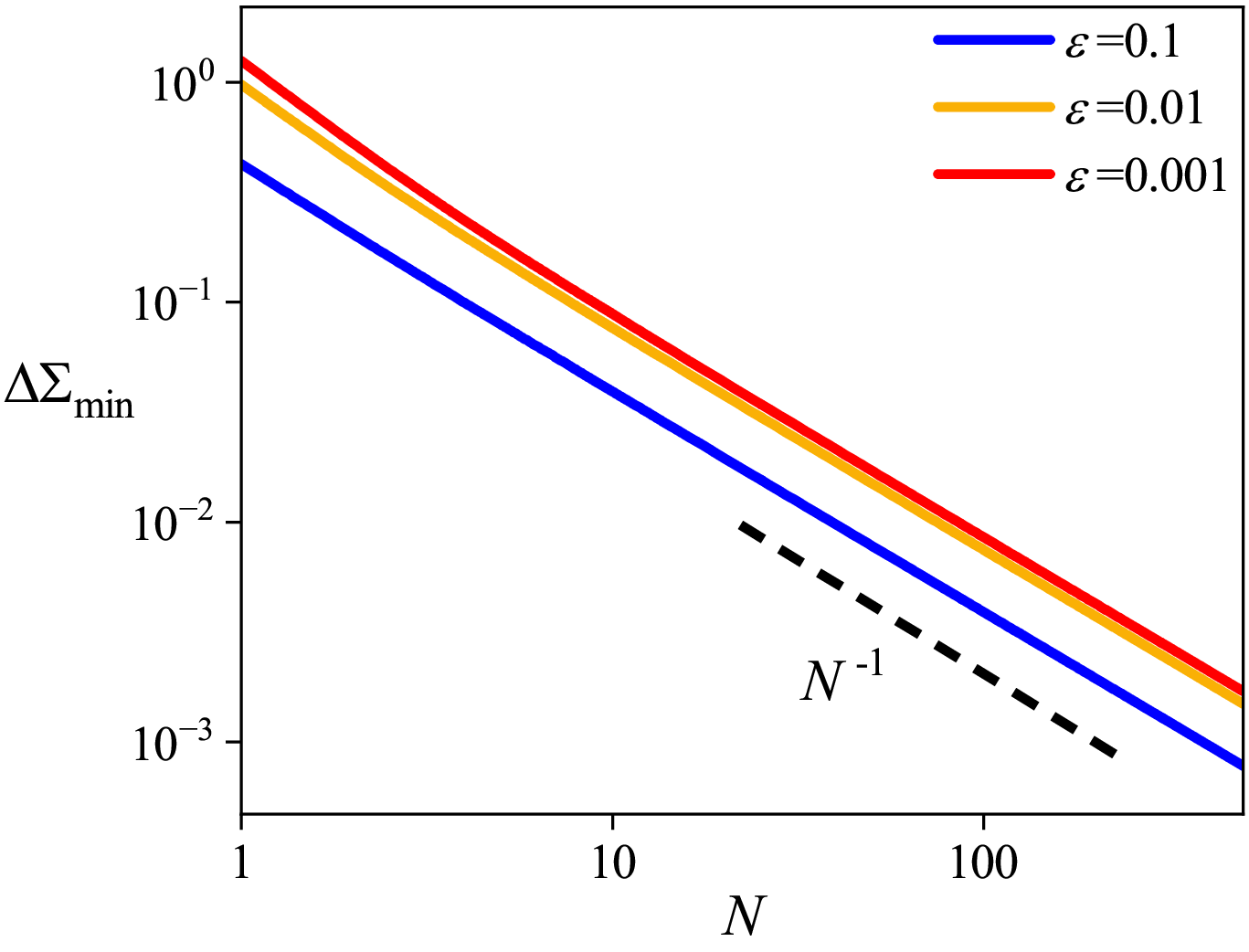}
\caption{Numerical results of the optimal protocol for the $2N$-state model. The minimal entropy production $\Delta\Sigma_{\text{min}}$ as a function of $N$ for different error probabilities $\varepsilon$ with the same erasure time $\tau=1$. The black dashed line guides the eye for the scaling of $N^{-1}$. Here, $\beta=1$ and $\mu=1$.}
\label{FigS3}
\end{figure}

Moreover, we also present the numerical results of the optimal protocol for different error probabilities $\varepsilon$. In Fig.~\ref{FigS3}, we show the minimal entropy production $\Delta\Sigma_{\text{min}}$ as a function of $N$ for different error probabilities $\varepsilon$ ($\varepsilon=0.001, 0.01~\text{and}~0.1$) with the same erasure time $\tau=1$. The numerical results show that the minimal entropy production $\Delta\Sigma_{\text{min}}$ approaches zero as $N^{-1}$ for different error probabilities $\varepsilon$.

\subsection{The collective spin model}
\begin{figure}[htbp]
\centering
\includegraphics[width=0.8\textwidth]{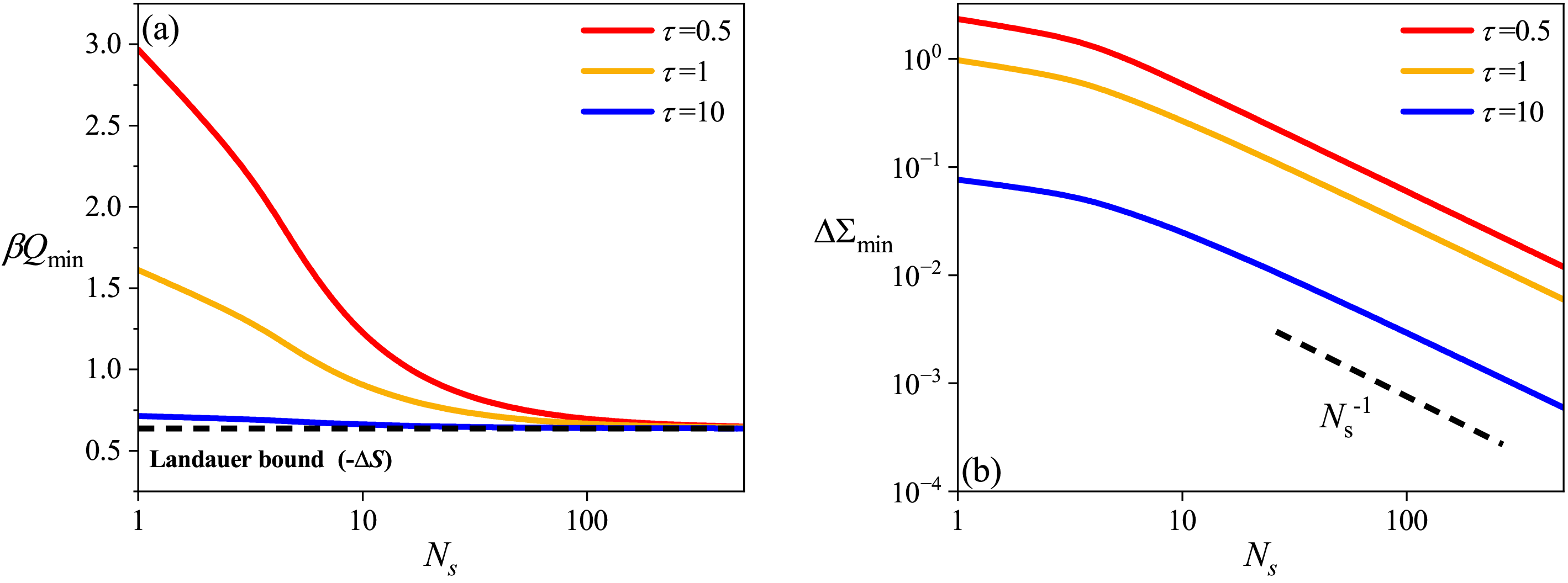}
\caption{Numerical results of the collective spin model. (a) The minimal heat dissipation $\beta Q_{\text{min}}$ as a function of $N_s$ for different erasure times $\tau$. (b) The minimal entropy production $\Delta\Sigma_{\text{min}}$ as a function of $N_s$ for different erasure times $\tau$. The black dashed line in (a) represents the Landauer bound $-\Delta S$, and the black dashed line in (b) guides the eye for the scaling of $N_s^{-1}$. Here, $\beta=1$, $\mu=1$ and $\varepsilon=0.01$.}
\label{FigS4}
\end{figure}

Furthermore, we also present similar numerical results of the collective spin model. In Fig.~\ref{FigS4}(a), we show the minimal heat dissipation $\beta Q_{\text{min}}$ as a function of the number of spins $N_s$ for different erasure times $\tau$. The numerical results show that the minimal heat dissipation $\beta Q_{\text{min}}$ approaches the Landauer bound $-\Delta S$ as $N_s$ increases for different erasure times $\tau$. In Fig.~\ref{FigS4}(b), we show the corresponding minimal entropy production $\Delta\Sigma_{\text{min}}$ as a function of the number of spins $N_s$ for different erasure times $\tau$. The numerical results show that the minimal entropy production $\Delta\Sigma_{\text{min}}$ approaches zero as $N_s^{-1}$ for different erasure times $\tau$. The numerical results of the collective spin model verify our analytical results.

\begin{figure}[htbp]
\centering
\includegraphics[width=0.8\textwidth]{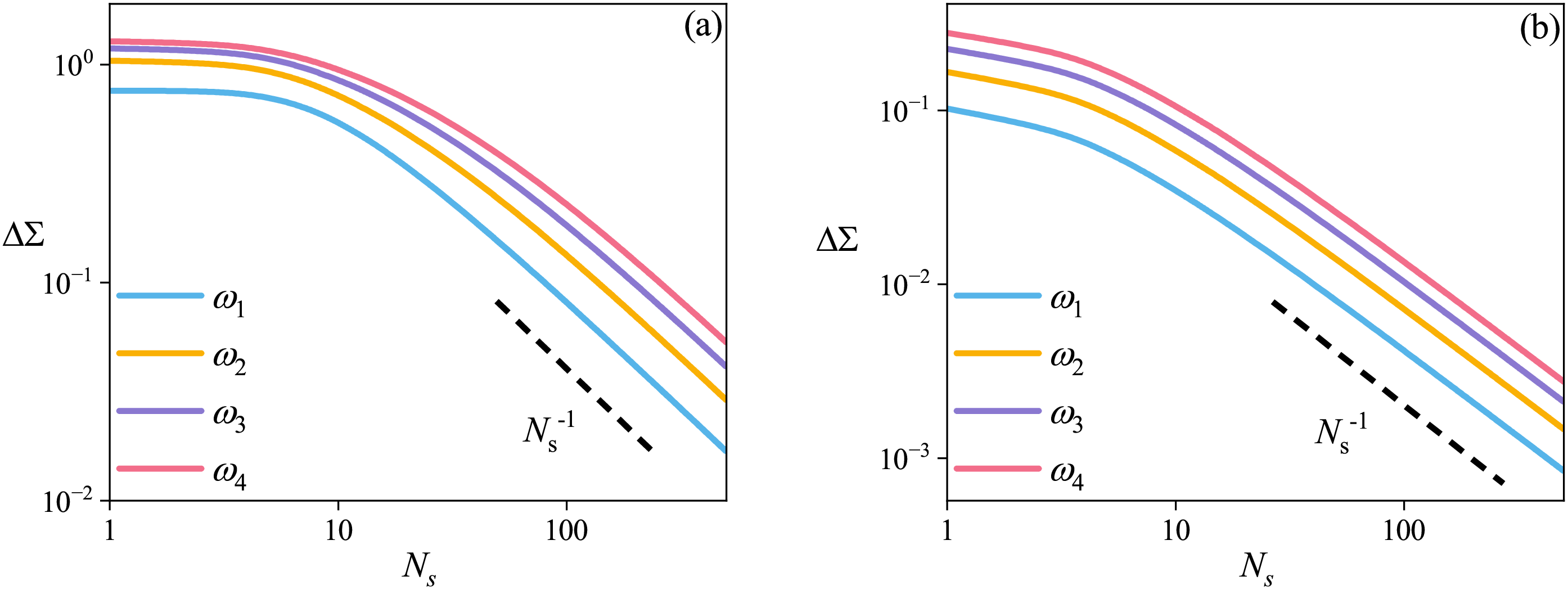}
\caption{Numerical results of the polynomial protocol for the collective spin model. The entropy production $\Delta\Sigma$ as a function of $N_s$ for four polynomial protocols with different values of $n$ ($n=1,2,3,4$) with the same erasure time $\tau=0.1$ in (a) and $\tau=10$ in (b). The black dashed lines guide the eye for the scaling of $N_s^{-1}$. Here, $\beta=1$ and $\mu=1$.}
\label{FigS5}
\end{figure}

Next, we also present the numerical results of the polynomial protocol for the collective spin model. We show the entropy production $\Delta\Sigma$ as a function of $N_s$ for four polynomial protocols with different values of $n$ ($n=1,2,3,4$) with the erasure time $\tau=0.1$ (Fig.~\ref{FigS5}(a)) and $\tau=10$ (Fig.~\ref{FigS5}(b)). The numerical results show that the entropy production $\Delta\Sigma$ approaches zero as $N_s^{-1}$ for polynomial protocols with different values of $n$.

\begin{figure}[htbp]
\centering
\includegraphics[width=0.4\textwidth]{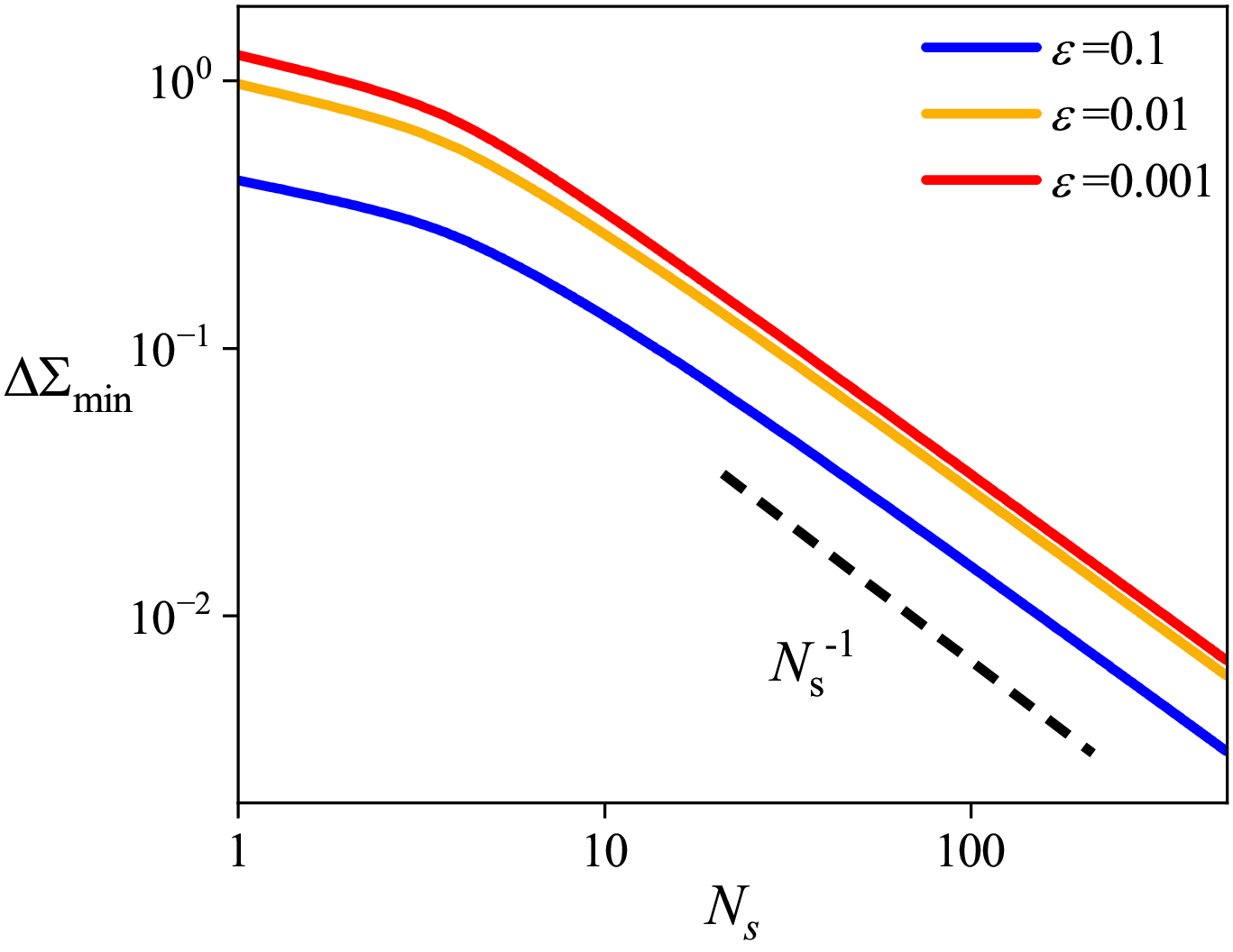}
\caption{Numerical results of the optimal protocol for the collective spin model. The minimal entropy production $\Delta\Sigma_{\text{min}}$ as a function of $N_s$ for different error probabilities $\varepsilon$ with the same erasure time $\tau=1$. The black dashed line guides the eye for the scaling of $N_s^{-1}$. Here, $\beta=1$ and $\mu=1$.}
\label{FigS6}
\end{figure}

Finally, we also present the numerical results of the optimal protocol for different error probabilities $\varepsilon$. In Fig.~\ref{FigS6}, we show the minimal entropy production $\Delta\Sigma_{\text{min}}$ as a function of $N_s$ for different error probabilities $\varepsilon$ ($\varepsilon=0.001,0.01~\text{and}~0.1$) with the same erasure time $\tau=1$. The numerical results show that the minimal entropy production $\Delta\Sigma_{\text{min}}$ approaches zero as $N_s^{-1}$ for different error probabilities $\varepsilon$.

